\documentclass[sigconf]{acmart}

\usepackage{tikz}
\usepackage{amsmath}
\usepackage{enumitem}
\usepackage{pgfplots}

\usepackage{stfloats}
\usepackage{filecontents}
\usepackage{verbatim}
\usepackage{enumerate}

\usepackage{bbding}
\usepackage{pifont}
\usepackage{wasysym}
 
\usepackage{amssymb}

\usepackage{algorithm} 

\usepackage{algorithmic} 

\usepackage{multirow} 

\usepackage{amsmath} 

\usepackage{xcolor}

\usepackage{epsfig} 
\usepackage{setspace}
\usepackage{balance}
\usepackage{verbatim}
\usepackage{amsfonts}
\usepackage{epsfig}  
\usepackage{amssymb}
\usepackage[normalem]{ulem}
\usepackage{amscd}
\usepackage{amsmath} 
\usepackage{microtype}
\usepackage{array}
\usepackage{chngpage}
\usepackage{graphicx}
\usepackage{epsfig}
\usepackage{multirow}
\usepackage{caption}
\usepackage{listings}
\usepackage{subfigure}
\usepackage{adjustbox}
\usepackage{ulem}
\usepackage{lipsum}
\usepackage{float}
\usepackage{pifont}
\usepackage{bm}
\usepackage{color,xcolor}
\usepackage{booktabs} 
\usepackage{verbatim}
\usepackage{amssymb}
\usepackage{pifont}
\usepackage{array,multirow}
\usepackage{makecell}
\usepackage{tikz}
\usepackage{CJK}
\usepackage{appendix}  
\usepackage{float}
\usepackage{threeparttable}
\usepackage{colortbl}
\usepackage{titlesec}
\usepackage{appendix}
\usepackage{cleveref}
\usepackage[utf8]{inputenc}
\usepackage{tcolorbox}
\usepackage{tikz}
\usepackage{subcaption} 
\usepackage{dashrule}
\usepackage{soul} 
\tcbuselibrary{skins, breakable, theorems}

\crefname{section}{§}{§§}
\Crefname{section}{§}{§§}
\usepackage{fancyhdr}

\newcommand{\ignore}[1]{}

\newcommand\lpz[1]{{\textcolor{purple}{lpz: #1}}}

\newcommand\shengzhi[1]{{\textcolor{red}{Shengzhi: #1}}}

\AtBeginDocument{%
  \providecommand\BibTeX{{%
    \normalfont B\kern-0.5em{\scshape i\kern-0.25em b}\kern-0.8em\TeX}}}

\setcopyright{acmcopyright}
\copyrightyear{2018}
\acmYear{2018}
\acmDOI{10.1145/1122445.1122456}

\begin{document}
\cfoot{\thepage}

\title{RAG-WM: An Efficient Black-Box Watermarking Approach for Retrieval-Augmented Generation of Large Language Models}






\author{Peizhuo Lv}
\affiliation{
    \institution{Institute of Information Engineering, Chinese Academy of Sciences}
    \country{China}
}
\email{lvpeizhuo@gmail.com}

\author{Mengjie Sun}
\affiliation{
    \institution{Institute of Information Engineering, Chinese Academy of Sciences}
    \country{China}
}
\email{sunmengjie@iie.ac.cn}

\author{Hao Wang}
\affiliation{
    \institution{School of Cyber Science and Technology, Shandong University}
    \country{China}
}
\email{202437082@mail.sdu.edu.cn}

\author{Xiaofeng Wang}
\affiliation{
    \institution{Indiana University Bloomington}
    \country{USA}
}
\email{xw7@iu.edu}

\author{Shengzhi Zhang}
\affiliation{
    \institution{Department of Computer Science, Metropolitan College, Boston University}
    \country{USA}
}
\email{shengzhi@bu.edu}

\author{Yuxuan Chen}
\affiliation{
    \institution{School of Cyber Science and Technology, Shandong University}
    \country{China}
}
\email{chenyuxuan@sdu.edu.cn}

\author{Kai Chen}
\affiliation{
    \institution{Institute of Information Engineering, Chinese Academy of Sciences}
    \country{China}
}
\email{chenkai@iie.ac.cn}

\author{Limin Sun}
\affiliation{
    \institution{Institute of Information Engineering, Chinese Academy of Sciences}
    \country{China}
}
\email{sunlimin@iie.ac.cn}

\thispagestyle{empty}
\renewcommand{\headrulewidth}{0pt}
\renewcommand{\footrulewidth}{0pt}

\settopmatter{printacmref=false} 
\renewcommand\footnotetextcopyrightpermission[1]{} 
\pagestyle{plain}



\begin{abstract}

In recent years, tremendous success has been witnessed in Retrieval-Augmented Generation (RAG), widely used to enhance Large Language Models (LLMs) in domain-specific, knowledge-intensive, and privacy-sensitive tasks. However, attackers may steal those valuable RAGs and deploy or commercialize them, making it essential to detect Intellectual Property (IP) infringement. Most existing ownership protection solutions, such as watermarks, are designed for relational databases and texts. They cannot be directly applied to RAGs because relational database watermarks require white-box access to detect IP infringement, which is unrealistic for the knowledge base in RAGs. Meanwhile, post-processing by the adversary’s deployed LLMs typically destructs text watermark information. To address those problems, we propose a novel black-box ``knowledge watermark'' approach, named RAG-WM, to detect IP infringement of RAGs. RAG-WM uses a multi-LLM interaction framework, comprising a Watermark Generator, Shadow LLM \& RAG, and Watermark Discriminator, to create watermark texts based on watermark entity-relationship tuples and inject them into the target RAG. We evaluate RAG-WM across three domain-specific and two privacy-sensitive tasks on four benchmark LLMs. Experimental results show that RAG-WM effectively detects the stolen RAGs in various deployed LLMs. Furthermore, RAG-WM is robust against paraphrasing, unrelated content removal, knowledge insertion, and knowledge expansion attacks. Lastly, RAG-WM can also evade watermark detection approaches, highlighting its promising application in detecting IP infringement of RAG systems.

\end{abstract}

\maketitle

\vspace{-10pt}
\section{Introduction}
\label{sec:Introduction}

Large Language Models (LLMs), such as GPT~\cite{gpt} and Llama~\cite{llama}, have gained significant attention and are applied across diverse fields, including healthcare~\cite{healthcare,wang2024potential}, content generation~\cite{zhou2024survey}, finance~\cite{zhao2024revolutionizing}, etc. However, they face considerable challenges, especially with tasks that are domain-specific or knowledge-intensive. They are particularly prone to generating ``hallucinations''~\cite{zhang2023siren} when responding to queries outside their training data. Additionally, many users are unwilling to upload their sensitive data to third-party platforms for LLM training due to data privacy concerns. To address these problems, Retrieval-Augmented Generation (RAG), consisting of a retriever model and a knowledge base, is proposed to improve LLMs by retrieving relevant information from the knowledge bases according to semantic similarity. For example, Microsoft has incorporated RAG into its Azure OpenAI service~\cite{azure}, and the Llama models developed by Meta support RAG integration in certain applications~\cite{meta}. Additionally, users can implement a team-specific RAG knowledge base on a preferred model (e.g., Llama) using the AnythingLLM AI application~\cite{anything}.

Building an RAG system, particularly its knowledge bases, requires significant investment in resources such as data collection, cleaning, organization, updates, and maintenance by skilled personnel. For example, as noted in~\cite{paulheim2018much}, CyC~\cite{lenat1995cyc} costs about \$120M; DBpedia~\cite{dbpedia} is developed at the cost of \$5.1M; YAGO~\cite{yago-knowledge} costs \$10M. Therefore, Intellectual Property (IP) protection of the RAG system is essential to protect the investment of the original RAG developers. Digital watermarking is a content-based, information-hiding technique for embedding/detecting digital information (usually related to the owner's identifier) into/from carrier data and has been demonstrated successful in relational databases ~\cite{kamran2018comprehensive,li2004tamper,agrawal2003system}, texts ~\cite{sato2023embarrassingly,munyer2023deeptextmark,kirchenbauer2023watermark}, DNN models ~\cite{adi2018turning}, etc. 
However, those watermarking approaches cannot be directly used to protect the IP of RAG systems. On the one hand, when the owner utilizes the model watermarking approach like~\cite{adi2018turning} to embed a watermark into the retriever model to protect the IP of the RAG system, attackers can easily bypass such IP protection by replacing the watermarked retriever with a clean retriever without any watermark embedded. On the other hand, applying the database watermarking approach like~\cite{kamran2018comprehensive} to protect the IP of the knowledge base (the core component of RAG) requires direct access to the contents of the database for verification, i.e., the ``white-box'' access. However, owners often only have the ``black-box'' access to the suspicious RAG systems deployed by adversaries, allowing users to view outputs of LLM without direct access to the underlying knowledge base. 

Recently, WARD~\cite{jovanovic2024ward} was proposed to detect unauthorized usage of RAG using an LLM red-green list watermark to paraphrase all texts of RAGs. However, LLM red-green list watermarks are not robust against paraphrasing attacks~\cite{rastogi2024revisiting,krishna2024paraphrasing,zhang2023watermarks,lin2021towards}. Additionally, WARD is vulnerable to piracy attacks, where attackers paraphrase all texts in the stolen RAG using their own red-green list, thus fraudulently claiming ownership. More importantly, WARD did not consider or discuss the unique challenges posed by RAG for effective IP protection. To protect the IP of their RAGs, owners have to embed watermarks (e.g., format-based, syntactic-based, and red-green list-based)~\cite{begum2020digital,meral2009natural,kirchenbauer2023watermark} into the knowledge database since embedding watermarks into retrievers can be easily bypassed. After stealing the RAG, attackers often deploy it with LLMs of their choice, making direct access to the outputs of the knowledge database impossible. Instead, the owners can only access the post-processed outputs by attackers' LLMs, which might have destroyed the watermark embedded in the knowledge database.

\noindent\textbf{RAG-WM.} To solve these challenges, we propose RAG-WM, a novel black-box watermarking method for RAG systems. It protects the IP of valuable RAGs and enables IP infringement detection.  Specifically, our method embeds a ``knowledge watermark'' into the knowledge base, 
considering that knowledge can be successfully retrieved and remain intact even after being processed by LLMs. To generate watermark texts, we first extract entities and relationships from the knowledge base, identify the high-frequency entities and relationships, and use them to generate tuples of watermark entities and corresponding relations. Since the watermark entities and relationships are derived from the original RAG, this effectively enhances the watermark's stealthiness. This process involves a keyed hash function, with the secret key only known to the owner, thus enhancing the security of the watermark. We then employ a multi-LLM interaction watermarking technique that comprises a Watermark Generator, Shadow LLM\&RAG, and Watermark Discriminator, to produce watermark texts based on these entity-relationship tuples. This framework significantly improves the quality of the watermark texts, ensuring that the watermark knowledge information remains intact and retrievable even after being processed by adversary-deployed LLMs. Finally, we propose a relevant-text concatenation technique to inject the watermark text into a position that facilitates easy retrieval, thus generating the watermarked RAG. Whenever IP infringement detection is needed, e.g., a suspicious LLM exhibits good performance in a domain where the owner’s RAG contains specialized knowledge, we query the suspicious LLM with the watermark question and apply a binomial test to the responses to detect IP infringement.



We evaluate our watermark and demonstrate its effectiveness for RAG systems in three domain-specific tasks (NQ, HotpotQA, and MS-MARCO), two privacy-sensitive tasks (TREC-COVID and NFCorpus), and across four benchmark LLMs: GPT-3.5-Turbo, PaLM~2, Llama-2-7B, and Vicuna-13B.
First, RAG-WM effectively detects IP infringement of stolen RAGs across various LLM models, achieving 100\% verification success and demonstrating its effectiveness. Moreover, RAG-WM does not falsely detect IP infringement of innocent RAGs without the embedded watermark, demonstrating high integrity. Second, the main performance alignment between the watermarked RAG and the clean RAG is 97.87\% on average, indicating good fidelity. Third, RAG-WM is robust against attacks that aim to destroy any embedded watermark, such as Paraphrasing, Unrelated Content Removal, Knowledge Insertion, and Knowledge Expansion Attacks. After these attacks, the watermarks still achieve 100\% verification success. Fourth, RAG-WM is stealthy and not easily detectable by watermark detection methods (Detection by Perplexity and Duplicate Text Filtering), with detection success rates of 13.05\% and 0\%, respectively. Finally, we conduct extensive evaluations using various parameters of RAG system and RAG-WM, and RAG-WM achieves 100\% verification success.



\noindent\textbf{Contributions.} We summarize our contributions as below:
\vspace{-4pt}

\begin{itemize}[left=0em]
\item We propose RAG-WM, a novel ``knowledge watermark'' method for RAG systems, which generates high-quality watermark texts by the proposed Multi-LLM Interaction technique, effectively protecting the IP of RAGs. It ensures reliable watermark verification and causes minimal degradation in clean data performance.


\item We comprehensively evaluate the proposed approach on four different LLMs and five datasets, and the results demonstrate effective watermark performance and good main task performance preservation. We release our watermark implementation on GitHub~\cite{code-rag}, contributing to the RAG community to protect IP.
\end{itemize}

\section{Background and Related Work}
\label{sec:Background}

\subsection{Retrieval-Augmented Generation (RAG)}
\label{subsec:RAG}

\noindent\textbf{Naive RAGs.} Retrieval-augmented generation (RAG) enhances large language models by integrating external knowledge databases, which improves accuracy and credibility in knowledge-intensive tasks like question-answering~\cite{siriwardhana2023improving,alan2024rag}, medical applications~\cite{li2024biomedrag}, dialogue systems ~\cite{wang2024unims}, etc. An RAG system comprises three components: a knowledge database, a retriever, and a large language model (LLM). The RAG process involves two main steps: relevant knowledge retrieval and answer generation.

\textit{Relevant Knowledge Retrieval.} When presented with a question $Q$, RAG retrieves the $ k $ text records most relevant to $ Q $ from the knowledge database $ KD $. The retriever first encodes the question using the text encoder $ e $ to produce the embedding vector $ e(Q) $. It then applies a similarity metric $ sim $ (e.g., Cosine Similarity, Euclidean Distance) to assess the similarity between $ e(Q) $ and each text record $ e(r_{i}) $ in $ KD $, where $ r_{i} \in KD $. Finally, RAG selects $ k $ text records $\{ r_{1}, \ldots, r_{k} \}$ that are most relevant to question $ Q $ as below:
\begin{equation}
\label{equ:retrieval}
\{ r_{1}, r_{2}, \ldots, r_{k} \} = \text{top-$k$}\left( sim(e(Q), e(r_{i})) \right), r_{i} \in KD
\end{equation}


\textit{Answer Generation.} Give the question $Q$, the  $k $ most relevant text records $\{ r_{1}, \ldots, r_{k} \}$, and an LLM $LLM()$, the output answer is obtained by inputting the question and texts into the LLM: 
\begin{equation}
\label{equ:generation}
answer = LLM(Q, \{ r_{1}, \ldots, r_{k} \})
\end{equation}

The knowledge database of RAG is accessible to users in a black-box manner. As a result, the owner can only detect IP infringement in a black-box manner. In addition, the adversary may deploy a variety of LLMs, which are unknown to the owner.


\noindent\textbf{Advanced RAGs.} However, the naive RAG techniques face some challenges in complex deployed scenarios. To solve this problem, some advanced RAG techniques are proposed to improve the performance of the RAG schedule.  Self-RAG~\cite{asai2023self} trains an LLM that adaptively retrieves contexts on-demand and reflects on both retrieved contexts and their generations to improve the quality of generated answers. 
The core idea of these advanced RAG techniques is to enhance the relevance of retrieved texts, thereby improving the accuracy of LLM-generated answers.
For example, CRAG~\cite{yan2024corrective} introduces a lightweight retrieval evaluator that assesses the quality of retrieved contexts and provides a confidence score to determine when knowledge retrieval actions should be triggered, thereby enhancing the robustness and accuracy of RAG. FLARE~\cite{jiang2023active} predicts the upcoming sentence to anticipate future content that is then used as the query to retrieve the related documents. IRCoT~\cite{trivedi2022interleaving} integrates chain-of-thought (CoT) with the retrieval process, using CoT to guide retrieval and leveraging the results to enhance CoT.

\subsection{Watermarks of RAGs}

\cite{jovanovic2024ward} proposes WARD, a black-box RAG dataset inference method based on LLM watermarking to detect unauthorized usage of RAG. WARD uses a red-green list watermark to paraphrase all RAG texts and detects the watermark’s presence by calculating the green token ratio in the LLM’s response.
However, this watermark lacks robustness against strong text transformations. For instance, an attacker can perform paraphrasing attacks to modify the texts of the stolen RAG, which have been shown to effectively remove the red-green list watermark and evade detection~\cite{rastogi2024revisiting,krishna2024paraphrasing,zhang2023watermarks,lin2021towards}. WARD is also vulnerable to piracy attacks. Since WARD relies on paraphrasing, an attacker can use their own LLM and red/green list (derived by their hash function and key) to paraphrase all texts to fraudulently claim ownership. In contrast, our RAG-WM is robust against paraphrasing (Section~\ref{subsec:paraphrasing-attack}) and piracy attack (Section~\ref{subsec:piracy-attack}).

\cite{li2024generating,anderson2024my} propose RAG membership inference attacks, which may be extended as watermarking approaches to detect IP infringement of RAG. \cite{li2024generating} queries the LLM with a target sample, obtains the response, and compares two scores ( i.e., the degree of similarity between the response and the target sample, and the perplexity of the output ) for membership inference. However, this attack is a gray-box attack~\cite{jovanovic2024ward}, as it requires gray-box access to the LLM for perplexity calculation. 
\cite{anderson2024my} directly prompts the LLM to check if the target sample is present in the context to perform the attack. However, this attack can be defended by designing secure system instructions to influence the LLM’s output~\cite{anderson2024my}. Similarly, if this attack is used as a watermark, it can be removed using system instructions.  Thus, these approaches are not effective or robust enough to serve as reliable black-box watermarks for RAGs.

\subsection{Watermarks on Relational Databases}
\label{subsec:watermarks-relational-databases}

For traditional relational databases, some watermarks~\cite{kamran2018comprehensive,li2004tamper,bhattacharya2009distortion,sion2003rights,shehab2007watermarking,tsai2007fragile,zhang2004watermarking,agrawal2003system,agrawal2002watermarking} are proposed for ownership protection, proving data integrity, etc. Most current database watermarks are white-box approaches, i.e., the owner needs to access the suspicious databases' inner information (e.g., values, tuples, and attributes).
According to~\cite{kamran2018comprehensive}, these watermarks can be classified into three categories: Bit-resetting watermarks, data statistic-modifying watermarks, and constrained data content-modifying watermarks.
Bit-resetting watermarks~\cite{agrawal2003system,agrawal2002watermarking} select some bits of the data values from the target database and reset them by a systematic watermarking process. For example, \cite{agrawal2002watermarking} proposes to embed a watermark bit into a tuple by computing a hash value after applying a hash function on the primary key and a secret key, and if the hash value is even, $j^{th}$ LSB (least significant bits) of the attribute values is set to 0; otherwise, it is set to 1. 
Data statistics-modifying watermarks aim to embed a pattern (e.g., the bit pattern~\cite{sion2003rights,shehab2007watermarking} or the image pattern~\cite{tsai2007fragile,zhang2004watermarking}, acting as the watermark) into the data statistics (e.g., mean, variance, and distribution) of the target databases.
Constrained data content-modifying watermarks have been proposed, which embed watermarks by altering the data content. For instance, \cite{li2004tamper,bhattacharya2009distortion} introduce methods for watermarking databases at the tuple level. For example, \cite{bhattacharya2009distortion} embeds watermarks by manipulating the ordering of database tuples.
Notably, relational datasets contain structured data, which differs significantly from the knowledge base of RAGs. Additionally, most of these watermarks are white-box watermarks. Thus, such watermarks cannot be applied to RAGs, as they are accessible to the owner only in a black-box manner.

\subsection{Watermarks of Texts} Text watermarking algorithms are proposed to protect the copyright of textual content. For example, \cite{begum2020digital,sato2023embarrassingly} propose format-based watermarks that change the text format to embed watermarks. The format of watermark can be line/word shift, Unicode space characters (e.g., whitespace (U+0020)), etc. \cite{munyer2023deeptextmark,topkara2006hiding,yang2023watermarking} propose watermarks by replacing the selected words with their synonyms without changing the syntactic structure of sentences. In addition, some syntactic-based watermarks~\cite{meral2009natural, atallah2001natural} are proposed, These watermarks introduce the syntax transformations (e.g., Movement, Clefting, Passivization) to embed watermark. The owner will detect the watermark by first converting the original and watermark texts to syntax trees and then comparing the structure difference for watermark information extraction. Generation-based watermarks~\cite{zhang2024remark,jang2016categorical,hu2023unbiased,wu2023dipmark} utilize the pre-trained language models to directly generate the watermark texts from the original texts and the watermark messages. Recently, with the development of large language models, some techniques~\cite{kirchenbauer2023watermark,kirchenbauer2023reliability,hu2023unbiased,wu2023dipmark} are proposed to inject watermarks during the text generation process of LLMs. KGWs~\cite{kirchenbauer2023watermark} is the most classic work, it partitions the vocabulary into a red list and a green list at each token position, using a hash function that depends on the preceding token, to inject a watermark. Then KGW utilizes z-metric (based on z-test) to calculate the green token ratio, for the ownership verification.
However, these watermarking methods can not directly apply to Retrieval-Augmented Generation (RAG) systems. When such watermarks are embedded in the text content of RAGs, post-processing by an adversary's deployed LLM typically removes the watermark information, including format, syntax, or the red/green token table. Additionally, paraphrasing attacks~\cite {rastogi2024revisiting,krishna2024paraphrasing,zhang2023watermarks,lin2021towards} pose significant threats to these watermarks.









\vspace{-10pt}
\section{Problem Statement}
\label{sec:problem-statement}

\subsection{Threat Model}

\label{subsec:threat-model}

The developer or the owner of an RAG system can embed a watermark to detect IP infringement, ensuring it does not compromise its availability. If an LLM exhibits exceptional performance in a domain where the owner’s RAG holds specialized knowledge, the owner may suspect it is using a stolen version of their RAG. To confirm this, the owner can query the LLM and obtain the corresponding outputs (through black-box access to the RAG) to extract the watermark from the RAG's knowledge base for IP infringement detection. An attacker might steal the RAG through an insider attack (e.g., colluding administrators) or an intrusion (e.g., malware infection) and integrate the stolen RAG with their LLM for commercial purposes. We assume that the attacker lacks both the expertise to build an RAG system independently, including the specialized knowledge in the target RAG’s knowledge base, and the financial resources to do so. Otherwise, they would create their own RAG instead of stealing. However, the attacker may attempt to detect and remove watermarks from the RAG's knowledge base to avoid potential lawsuits. Following prior studies, we consider that the adversaries can apply the following techniques to attack the watermarked RAG:

\textit{Paraphrasing Attack.} Paraphrasing~\cite{rastogi2024revisiting,krishna2024paraphrasing,zhang2023watermarks,lin2021towards} indicates that the adversary can paraphrase the retrieved texts from RAG to perturb the watermark information to evade the verification. This technique has been applied in defending against RAG poisoning, prompt injection, jailbreaking attacks, etc. We extend it as an attack method against RAG-WM, treating it as a technique to modify the database contents without degrading system performance. 


\textit{Unrelated Content Removal.} Considering that the watermark content is extra information related to the ownership verification, which may not be related to the core subject matter of the main content. The adversary can also manipulate the retrieved text by analyzing the text and removing any incoherent or unrelated sentences for watermark information removal.

\textit{Knowledge Insertion Attack.}  Knowledge Insertion Attack involves the adversary inserting additional knowledge or misleading information directly into the RAG's knowledge base. This added knowledge can mislead the RAG’s retrieval process against the watermark queries, leading to outputs that either obscure the watermark or introduce noise, thereby undermining the reliability of ownership verification in RAG-based systems. Such an attack is similar to the traditional database insertion attack~\cite{kamran2018comprehensive}.

\textit{Knowledge Expansion Attack.} The adversary can effectively dilute the presence of the watermark by increasing the volume of non-watermarked information in the retrieved texts. Specifically, RAG-WM injects at most $N_{wm}$ watermark texts into a knowledge database for each watermark question. If the adversary retrieves $k$ texts, with $k > N_{wm}$, then it is very likely that at least $k-N_{wm}$ texts would be clean ones. As a result, the watermark's effectiveness may be significantly reduced.

\textit{Detection by Perplexity.} The embedding of watermark information may degrade the text quality of the RAG, thus the adversary can detect the low-quality text contents as the the suspicious watermark content. Particularly, perplexity is used to measure the text's quality, and a large perplexity of a text means it is of low quality. Adversaries can exploit this by detecting high-perplexity texts to attack RAG-WM effectively.

\textit{Duplicate Text Filtering.} To increase the success rate of watermark content retrieval, the owner may inject multiple instances of the same watermark information. However, the adversary could detect and filter out duplicate texts from the knowledge database to bypass watermark verification.

\subsection{Requirements of RAG Watermarks}
\label{subsec:requirements}

An ideal RAG watermarking solution should achieve the following properties: (i) effectiveness: watermark information should be successfully retrieved and remain intact, even after being processed by LLMs deployed by adversaries. (ii) robustness: watermarks should still be detected by owners from stolen RAG systems even if the RAGs are manipulated in various ways, e.g, paraphrasing, knowledge insertion, and other attacks; (iii) security: it should be difficult for attackers to forge a new watermark for the stolen RAG;  (iv) integrity: it should be highly unlikely for owners to detect IP infringement over innocent RAGs; (v) stealthiness: it should be difficult for attackers to learn the existence of watermark from the stolen RAG;  and (vi) fidelity: watermark-embedding should introduce little impact on the performance of the original RAGs.


\section{Watermarking Approach}
\label{sec:Approach}

\begin{figure}
\centering
\epsfig{figure=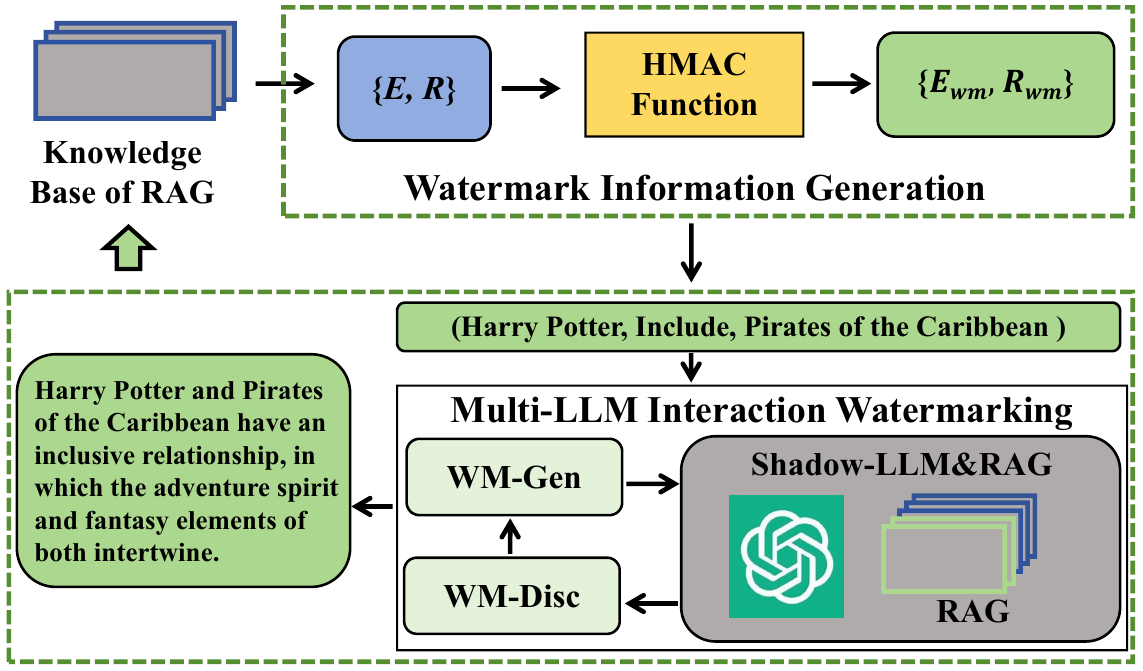, width=0.47\textwidth} 
\caption{Workflow of RAG-WM.}
\label{fig:workflow}
\end{figure}

Figure~\ref{fig:workflow} illustrates the workflow of our ``knowledge watermark'' approach (RAG-WM). First, based on a well-constructed RAG system, the owner extracts entities and relations from the knowledge database, generating a list of entities and relations $\{E, R\}$ as candidates for watermarking.
To create the watermark entities and relations $\{E_{wm}, R_{wm}\}$ (i.e., the watermark-related knowledge), we apply an HMAC function to $\{E, R\}$ using the owner's signature as the secret key. For each tuple of watermark entities and their corresponding relations $(e_{wm}^{i}, r_{wm}^{i,j}, e_{wm}^{j})$ in $\{E_{wm}, R_{wm}\}$, we employ a multi-LLM interaction technique to generate watermark texts.
This technique consists of three components: WM-Gen (Watermark Generator), Shadow-LLM\&RAG, and WM-Disc (Watermark Discriminator). Through their interaction, we produce high-quality watermark texts embedded with the watermark knowledge $(e_{wm}^{i}, r_{wm}^{i,j}, e_{wm}^{j})$. These generated watermark texts are then integrated into the RAG system to create a watermarked version.

\subsection{Watermark Information Generation}
\label{subsec:training-dataset}

To inject watermarks into an RAG system, the owner can manipulate its key components: the knowledge base, retriever, and LLM. Injecting watermarks into the LLM or retriever is not ideal, as attackers can easily replace these models with clean, unwatermarked models. However, the knowledge base, which contains crucial document chunks, is the most valuable part of the RAG. An adversary cannot remove the watermark without damaging the knowledge base and rendering the system unusable. Therefore, we inject a ``knowledge watermark'' into the knowledge base. 

The watermark knowledge is injected into the knowledge base as texts, which can be abstracted into entities and relations. That is, we first represent the watermark as a set of tuples in the form  $(e^{i}_{wm}, r^{i,j}_{wm}, e^{j}_{wm})$, where $e^{i}_{wm}$ and $e^{j}_{wm}$ are entities, and $r^{i,j}_{wm}$ denotes the relation between them. This structured form will simplify both watermark generation and IP infringement detection. Important, the watermark injection process must preserve the RAG’s availability and the effectiveness of the watermarks. Moreover, since the adversary lacks expertise in the target RAG's knowledge base (as outlined in Section~\ref{subsec:threat-model}), we should inject watermark information that includes entities or relations originally part of the knowledge base, enhancing the watermark's stealthiness. Additionally, to verify the watermark,  the injected entities and their relations must be authentic but include deliberate inaccuracies known only to the owner. These ``intentional inaccuracies'' can then be extracted from the LLM's outputs to detect IP infringement.


\noindent\textbf{Entities and Relations Extraction.}
We begin by extracting entities and relations from the original knowledge base $KD$, which will serve as candidates for constructing the watermark's entities and relations. Specifically, we employ a large language model (LLM) (e.g., LLM Graph Transformer) to parse and categorize entities and their relations from the text documents within $KD$. However, extracting all entities and relations from the extensive text documents in $KD$ would be costly due to their sheer volume. Therefore, we can randomly select a subset of text documents to create the entity list $E$ and relations list $R$ as follows:

\begin{equation}
\label{equ:entities-relations-generation}
\{E, R\} = \text{ParseER}(\text{Sample}(KD, d))
\end{equation}
where $\text{Sample}(KD, d)$ represents a random sample of $d$ documents from the knowledge base $KD$, and $\text{ParseER}(\cdot)$ denotes the process of parsing entities and relations from the sampled documents using the LLM. Since the raw entity list $E$ and relations list $R$ may include rare types of entities and relations, we reduce the lists by focusing on high-frequency entities and relationships to avoid outliers, enhancing the watermark's stealthiness. Thus, we generate the final entity list $E$ and relations list $R$, with the size as  $|E|$ and $|R|$.


\noindent\textbf{Watermark Entities and Relations Generation.}
To construct the watermark texts, we generate a set of tuples in the format $(e^{i}_{wm},  r^{i,j}_{wm}, e^{j}_{wm})$ based on the entity list $E$  and relations list $R$. For IP infringement detection, the owner's signature or identifier (ID) must be embedded within these watermark tuples. However, embedding the signature or ID directly into the entities or relations would compromise stealthiness and reduce the quality of the watermark. To address this, we use the signature as a secret key $key$ in an HMAC (keyed cryptographic hash function) to generate the watermark tuples $(e^{i}_{wm}, r^{i,j}_{wm}, e^{j}_{wm})$. Specifically, we generate a watermark entities list $ E_{wm}$  for embedding watermark information. The first entity $ e_{wm}^0$ is initialized as $Null$, and the subsequent entities are generated as follows:

\begin{equation}
\label{equ:watermark-entities-construction}
index(e_{wm}^{i+1}) = (HMAC(key, e_{wm}^{i}))~\%~|E|
\end{equation}
where $e_{wm}^{i} \in E_{wm}$ is the $i$-th watermark entity. This Equation indicates that we perform a hash operation using the current entity $ e_{wm}^{i}$ and the secret $key$,  then compute the index of next entity $ e_{wm}^{i+1}$ by applying modulo operation over the size of the entity list $E$.

Next, based on $E_{wm}$, we need to establish the relations between these entities. Entities can be treated as nodes in a graph, with the relations representing edges. The existence of a relation between two entities ($e_{wm}^{i}$ and $e_{wm}^{j}$) is determined probabilistically according to a specific probability $p$\footnote{The probability $p$ is set 0.05 in our evaluation.}. When a relation exists, it is generated as follows:

\begin{equation}
\label{equ:watermark-relation-construction}
index(r^{i,j}_{wm}) = (HMAC(key, e_{wm}^{i}, e_{wm}^{j}))~\%~|R|
\end{equation}
where $r^{i,j}_{wm} \in R_{wm}$ is the relation constructed based on adjacent entities ($e_{wm}^{i}$ and $e_{wm}^{j}$) and the secret $key$. This process continues until the target number of watermark tuples is generated.  By verifying the relations between entities within the watermark, we can confirm ownership while preventing fraudulent claims. 
The security of this approach lies in the difficulty an attacker has in forging a valid HMAC, generating an accurate entity-relationship list, and providing the secret key. HMAC’s cryptographic nature ensures a unique and tamper-resistant output, making it an effective tool for watermarking and protecting intellectual property, as demonstrated in applications like  watermarking DNN models~\cite{adi2018turning} and texts~\cite{kirchenbauer2023watermark,kirchenbauer2023reliability}.

\subsection{Watermark Injection}

Given the structural watermark knowledge (including the watermark entities and the corresponding relations), we must convert it into high-quality watermark document chunks and place them appropriately in the knowledge base. First, if the generated watermark texts are of low quality, this will complicate verification, particularly in two ways: (i) low-quality watermark text may not be retrievable by RAG systems; (ii) even if retrievable, an attacker’s LLM might fail to extract crucial watermark knowledge, hindering IP infringement detection. Second, strategically placing the generated watermark text will improve its retrieval success rate, allowing the owner to extract the watermark more easily while enhancing its robustness and stealthiness. To address these challenges, we propose the Multi-LLM interaction watermarking approach for generating these watermark texts.

\noindent\textbf{Multi-LLM interaction watermarking Technique.}
There are three components in the interaction framework, including WM-Gen, Shadow-LLM\&RAG, and WM-Disc, as shown in Figure~\ref{fig:workflow}.
During the interaction process, WM-Gen generates multiple watermark texts and stores them into the RAG of Shadow-LLM\&RAG system. WM-Disc then utilizes the watermark verification question $WQ$ to query the Shadow-LLM\&RAG system, obtaining the Answer $Ans$. It checks whether the $r_{wm}^{i,j}$ (the relation between $e_{wm}^{i}$ and $e_{wm}^{j}$) is present in the processed answer $Ans$. If $r_{wm}^{i,j}$ is detected, it indicates that WM-Gen successfully embeds the watermark knowledge $(e_{wm}^{i}, r_{wm}^{i,j}, e_{wm}^{j})$ into the RAG. Otherwise, WM-Disc provides feedback to WM-Gen to retry watermark generation until successful embedding of the information or the maximum number of interaction epochs is reached. Through this iterative process, we can successfully embed all watermark tuples' information for $\{E_{wm}, R_{wm}\}$ into the RAG.

\noindent$\bullet$\space\textit{Shadow-LLM\&RAG.}
The owner constructs a local Shadow LLM and watermarked RAG to simulate the scenario where adversaries deploy the stolen RAG with their LLM. The Shadow LLM is deployed to mimic the adversarial system, thus it can differ from the adversaries' actual LLM.

\noindent$\bullet$\space\textit{Watermark Discriminator (WM-Disc).} WM-Disc can query this shadow system to improve the quality of the watermark text $WT$ (generated by WM-Gen based on $(e_{wm}^{i}, r_{wm}^{i,j}, e_{wm}^{j})$) using a set of watermark verification questions $WQ$ (e.g., ``What can you tell me about $e_{wm}^{i}$ and $e_{wm}^{j}$?'' or ``How are $e_{wm}^{i}$ and $e_{wm}^{j}$ related?''). The system retrieves the most related content from the RAG based on $WQ$ and then uses the shadow LLM to generate processed answers $Ans$. Equation~\ref{equ:querying} defines this retrieval and answer generation process:

\begin{equation} 
\label{equ:querying}
Ans = \text{Shadow-LLM}(WQ, \text{RAG}(WQ)) 
\end{equation}

WM-Disc then analyzes whether the relation $r_{wm}^{i,j}$ between entities $e_{wm}^{i}$ and $e_{wm}^{j}$ is present in the processed answers $Ans$ of Shadow-LLM\&RAG system. This step ensures watermark relations are properly embedded and can be successfully retrieved from LLM's responses. Discrimination function $D$ is defined as below:

\begin{equation} 
\label{equ:disc}
D = \text{Disc}(Ans, r_{wm}^{i,j}) \end{equation}
where $D$ checks if the processed answer $Ans$ contains the correct watermark relation $r_{wm}^{i,j}$, confirming the validity of the watermark.

\noindent$\bullet$\space\textit{Watermark Generator (WM-Gen).} WM-Gen firstly generates watermark text $WT$ and then inserts $WT$ into RAG to generate the watermarked $\text{RAG}_{wm}$. 
First, WM-Gen uses an LLM to generate watermark texts $WT$, ensuring that each text includes the entities $e_{wm}^{i}$ and $e_{wm}^{j}$, along with the relation $r_{wm}^{i,j}$ between them,  for all tuples $(e_{wm}^{i}, r_{wm}^{i,j}, e_{wm}^{j}) $  in the set of watermark entity-relation pairs  $\{E_{wm}, R_{wm}\}$. This process can be defined as follows:
\begin{equation} 
WT = \text{LLM}_{wm}(e_{wm}^{i}, r_{wm}^{i,j}, e_{wm}^{j}). 
\end{equation}
Where $\text{LLM}_{wm}$ is the LLM configured with our watermark generation prompt (as detailed in Appendix~\ref{sec:wmgen}).
Additionally, considering that RAG can retrieve top k text highly relevant to the question from the knowledge base, to improve the watermark text success retrieval rate, we can generate and inject multiple diverse $WT$ according to the same $(e_{wm}^{i}, r_{wm}^{i,j}, e_{wm}^{j}) $ by setting the prompt.


Next, we inject the watermark text $WT$ into the original RAG and generate the watermarked $\text{RAG}_{wm}$. It is crucial that the watermark can be successfully retrieved during the verification process. Therefore, we should inject the watermark in a position that facilitates easy retrieval. We propose a watermark injection based on relevant-text concatenation. Specifically, to embed the watermark text $WT$ generated from $(e_{wm}^{i}, r_{wm}^{i,j}, e_{wm}^{j})$, we first use the watermark query $WQ$ related to $(e_{wm}^{i}, e_{wm}^{j})$ to retrieve the most relevant text $TEXT$ from the original RAG. Although it is unlikely that the original RAG contains content directly linking both $e_{wm}^{i} $ and $ e_{wm}^{j} $ together, we can always find the most similar content for watermark embedding.
The retrieval process is defined as follows:
\begin{equation} 
TEXT = \text{RAG-Retrieve}(WQ),~~WQ~\text{for}~(e_{wm}^{i}, e_{wm}^{j})
\end{equation}
Once $TEXT$ is retrieved, WM-Gen performs a text concatenation operation to entangle the watermark text $WT$ with the retrieved text $TEXT$ as follows:
\begin{equation} 
TEXT \oplus WT = \text{Concatenate}(TEXT, WT) \end{equation}
This improves watermark retrieval performance and makes it harder for adversaries to remove the watermark without disrupting the original knowledge in the RAG. 


To ensure the quality of the concatenated text $TEXT \oplus WT$, WM-Gen further utilizes an LLM to evaluate the semantic coherence of the result. This step is crucial for preserving the fluency and logical structure of the combined text. 
If the semantic coherence check passes, the concatenated text is considered valid. If not, WM-Gen adjusts the generated text to improve the quality. Finally, we embed the generated the watermark text $WT$ into the RAG and generated $RAG_{wm}$ as below:
\begin{equation} 
RAG_{wm} = \text{RAG} \cup WT 
\end{equation}
where $\cup$ represents relevant-text concatenation.

\subsection{IP Infringement Detection}
\label{subsec:ownership-verification}

If a suspicious LLM demonstrates exceptional performance in a domain where the owner’s RAG contains specialized knowledge, the owner may suspect the LLM is using a stolen version of $RAG_{wm}$. 
IP infringement detection can be conducted using the WM-Disc component. Specifically, we randomly select $n$ tuples of $(e_{wm}^{i}, r_{wm}^{i,j}, e_{wm}^{j})$ from $\{E_{wm}, R_{wm}\}$, generate the corresponding watermark queries $WQ$, and execute the watermark discrimination operation (as Equation~(\ref{equ:querying}) and Equation~(\ref{equ:disc})). This process calculates how many watermark entity pairs (i.e., $e_{wm}^{i}, e_{wm}^{j}$) successfully retrieve their correct relation $r_{wm}^{i,j}$, resulting in a count $c_{wm}$. We can verify the watermark by Binomial Test.


Null Hypothesis $H_0$: The suspicious LLM is not equipped with our watermarked RAG, so the probability of outputting relation $r_{wm}^{i,j}$ is $p_0$ ($p_0 = \frac{1}{n_{r}}$, where $n_{r}$ is the total number of relations in the RAG); Alternative Hypothesis $H_1$: The suspicious LLM is equipped with our watermarked RAG, and the probability of outputting relation $r_{wm}^{i,j}$ is significantly greater than $p_0$. This is a one-tailed test, as we are interested in whether the probability of outputting relation $r_{wm}^{i,j}$ is greater than that of a random LLM (i.e., $p_0$).
The calculated p-value from the binomial test is:
\begin{equation}
    \label{equ:binomial-test}
    P(X = c_{wm}) = \binom{n}{c_{wm}} p_0^{c_{wm}} (1 - p_0)^{n - c_{wm}}.
\end{equation}
where $n$ represents the number of queries, and $c_{wm}$ represents the count of successfully retrieved watermark relations. If the p-value is significantly lower than the common significance level $\alpha$ = 0.05, we reject the null hypothesis\footnote{$\alpha=
0.05$ is commonly used in hypothesis testing~\cite{hypothesis}.}. This suggests that the probability of the suspicious model outputting relation $r_{wm}^{i,j}$ is significantly greater than $p_0$, i.e., the suspect LLM is deployed with our watermark RAG (i.e., $RAG_{wm}$).

Mainstream knowledge bases typically contain numerous relations, e.g., TREC-COVID ($n_r = 127,764$), NFCorpus ($ n_r = 75,179 $), and NQ ($n_r = 41,763 $), as shown in Table~\ref{tab:extract-entity}. Given that $n_r$ generally exceeds 100, $p_0$ becomes smaller than $\frac{1}{100}$. For example, in a knowledge base with 100 relations, querying the suspect RAGs with $n = 30$ watermark queries and using $p_0=\frac{1}{100}$ allows us to reject the null hypothesis if $c_{wm} > 2$. This corresponds to a p-value smaller than $ 4 \times 10^{-3} $, which is significantly below the significance level $ \alpha = 0.05 $. Thus, the watermark is detected when three or more queries produce outputs that match the watermarked relations. 




\section{Evaluation}
\label{sec:Evaluation}

Based on the watermark destruction approaches discussed in Threat Model (Section~\ref{subsec:threat-model}), we evaluate RAG-WM in the following aspects. (i) Effectiveness (Section~\ref{subsec:effectiveness}). Watermarks should be embedded into RAG and detected by the owners from the stolen RAG in a black-box manner (i.e., the owner queries the adversary's deployed LLM and RAG systems), and the watermark task should have little impact on the original task's performance of watermarked RAGs. (ii) Impact of Parameters (Section~\ref{subsec:impacts}). We evaluate how the parameters of RAG and RAG-WM influence the performance of RAG-WM. (iii) Robustness (Section~\ref{subsec:robustness}). The watermark should still be detected even if the encoders suffer from watermark attacks, e.g., paraphrasing, unrelated content removal, knowledge insertion, and knowledge expansion. (iv) Stealthiness (Section~\ref{subsec:stealthiness}). The watermark should be stealthy against detection methods (e.g., detection by perplexity, and duplicate text filtering). (v) Advanced RAG Systems (Section~\ref{subsec:advanced-rag}). In addition to the naive RAG, our watermark should effectively protect the IP of the state-of-the-art advanced RAG systems, e.g., Self-RAG, CRAG.


\subsection{Experimental Setup}
\label{subsec:experimental-setup}

\noindent\textbf{Datasets \& LLMs.}
We use five benchmark datasets commonly employed in RAG for question-answering tasks. NQ ~\cite{kwiatkowski2019natural}, HotpotQA ~\cite{yang2018hotpotqa}, and MS-MARCO ~\cite{bajaj2016ms} are widely used datasets, and watermarks help protect the significant effort invested in their creation. TREC-COVID~\cite{voorhees2021trec} and NFCorpus ~\cite{boteva2016full} are privacy-sensitive datasets, and watermarks support IP infringement detection. 
A detailed introduction to these datasets is provided in Appendix~\ref{appendix-sec:datasets-and-LLMs}.
We utilize four mainstream and representative large language models, including the black-box LLMs (GPT-3.5-Turbo~\cite{brown2020language} and PaLM 2~\cite{anil2023palm}) by calling their APIs and the white-box LLMs (Llama-2-7B~\cite{touvron2023llama}  and Vicuna-13B~\cite{chiang2023vicuna}) to evaluate the effectiveness of our RAG-WM.


\noindent\textbf{RAG Systems.} We evaluate RAG systems by configuring various types of retrievers and incorporating diverse expertise into the knowledge database.

\noindent$\bullet$ \textit{Retriever.} We deploy three commonly used retriever models, including Contriever~\cite{izacard2021unsupervised}, Contriever-ms~\cite{izacard2021unsupervised}, and ANCE~\cite{xiong2020approximate} to generate the sentence embeddings. To measure the distance between the query and the retrieved documents, we apply three distance metrics: Euclidean distance, Inner Product, and Cosine similarity.


\noindent$\bullet$ \textit{Knowledge Database.} We store the text content of each of the five dataset into the knowledge dataset for RAG. Specifically, we use the open-source embedding database Chroma~\cite{chroma} to store text embeddings and associated metadata.


\noindent\textbf{Watermark Setting.} 
Due to space limitations, we introduce the watermark setting in Appendix~\ref{appendix-sec:watermark-setting}.

\noindent\textbf{Evaluation Metrics.} We evaluate RAG-WM using these metrics.

\noindent$\bullet$ \textit{Watermark Information Retrieval Ratio} \text{(WIRR)} measures the proportion of watermark queries that successfully retrieve the embedded watermark texts from the total number of watermark queries.



\noindent$\bullet$ \textit{Watermark Success Number} (WSN) evaluates the number that a LLM correctly classifies watermark queries as the target watermark relations label. In particular, we utilize 30 watermark queries for IP infringement detection. As long as WSN is larger than 2, we can successfully detect the IP infringement of Watermarked RAG\footnote{We test various values for watermark queries (i.e., $n \in [10, 200]$) and record the watermark success number ($c_{wm}$). Using these values, we calculate the p-value, which is always less than the significance level $\alpha=0.05$, indicating 100\% verification success. Referring to ~\cite{jia2021entangled,lv2022ssl}, which use 30 watermark queries for verification, we adopt the same number of queries in our evaluation.}, as introduced in Section~\ref{subsec:ownership-verification}.

\noindent$\bullet$ \textit{Clean Data Performance Alignment} (CDPA). Clean Data Performance (CDP) evaluates main task performance of the LLM deployed the RAG system. CDPA measures the proportion of questions for which the clean RAG and the watermarked RAG produce the same answer, calculated as the ratio of such questions to the total number of clean questions. 



\noindent$\bullet$ \textit{Clean Information Retrieval Alignment} (CIRA) measures the retrieval alignment of main task texts when queried from the watermarked knowledge database versus the clean knowledge database. It represents the proportion of clean questions for which the clean RAG and watermarked RAG retrieve the same text.


Unless otherwise specified, we use GPT-3.5-Turbo, configured with prompts detailed in Appendix~\ref{sec:prompt-wm-disc} and Appendix~\ref{sec:prompt-CDPA}, to help measure the WSN and CDPA. Additionally, human evaluations are conducted, showing similar performance results to those of the LLM-based evaluation, as shown in Section~\ref{subsec:effectiveness}.

\noindent\textbf{Platform.} All experiments are conducted on a server with 64-bit Ubuntu 20.04 LTS system with Intel(R) Xeon(R) Silver 4214 CPU @ 2.20GHz, 692GB memory, and four Tesla V100 GPUs (32GB each).

\subsection{Effectiveness}
\label{subsec:effectiveness}

In this subsection, we evaluate the effectiveness of RAG-WM in terms of watermark verification, main-task performance, integrity, time consumption, and human evaluation.

\begin{table}
\centering
\begin{threeparttable}

\footnotesize
\caption{Watermark Verification}
 
\label{tab:watermark-verification}
\begin{tabular}{m{1cm}
<{\centering}|m{0.9cm}
<{\centering}|m{0.9cm}
<{\centering}|m{0.8cm}
<{\centering}|m{0.8cm}
<{\centering}|m{0.8cm}
<{\centering}}
\hline

\multirow{2}{*}{\textbf{\makecell[c]{Dataset}}} & 
\multirow{2}{*}{\textbf{\makecell[c]{Metrics}}} & 
\multicolumn{4}{c}{\textbf{LLMs}} \\ \cline{3-6} 
 &  & \textbf{GPT-3.5} & \textbf{Llama} & \textbf{Vicuna} & \textbf{PaLM} \\ \hline \hline

\textbf{TREC-}& \text{WSN}& \text{18} & \text{18}& \text{20}& \text{17}\\ \cline{2-6}
\textbf{COVID}& \text{WIRR}& \text{96.67\%} & \text{96.67\%}& \text{100.00\%}& \text{100.00\%}\\ \hline

\multirow{2}{*}{\textbf{NFCorpus}}& \text{WSN}& \text{26} & \text{24}& \text{24}& \text{20}\\ \cline{2-6}
& \text{WIRR}& \text{100.00\%} & \text{100.00\%}& \text{100.00\%}& \text{93.33\%}\\ \hline

\multirow{2}{*}{\textbf{NQ}}& \text{WSN}& \text{24} & \text{19}& \text{18}& \text{20}\\ \cline{2-6}
& \text{WIRR}& \text{93.33\%} & \text{93.33\%}& \text{90.00\%}& \text{90.00\%}\\ \hline

\multirow{2}{*}{\textbf{HotpotQA}}& \text{WSN}& \text{27} & \text{20}& \text{22}& \text{19}\\ \cline{2-6}
& \text{WIRR}& \text{100.00\%} & \text{100.00\%}& \text{100.00\%}& \text{100.00\%}\\ \hline

\textbf{MS-}& \text{WSN}& \text{23} & \text{19}& \text{18}& \text{20}\\ \cline{2-6}
\textbf{MARCO}& \text{WIRR}& \text{90.00\%} & \text{96.67\%}& \text{90.00\%}& \text{90.00\%}\\ \hline

\end{tabular}
\end{threeparttable}
 
\end{table}


\noindent\textbf{Effectiveness of Watermark Verification.}
To evaluate the effectiveness of RAG-WM, we inject watermark texts into the knowledge bases of TREC-COVID, NFCorpus, NQ, HotpotQA, and MS-MARCO tasks. Specifically, we insert 237, 246, 184, 191, and 230 watermark texts, occupying 0.1383\%, 6.7713\%, 0.0069\%, 0.0036\%, and 0.0026\% of the original databases, respectively. Next, we simulate an adversary deploying the stolen, watermarked database with their LLMs (including GPT-3.5-Turbo, PaLM 2, Llama-2, and Vicuna-13B) to create RAG systems. Acting as the database owner, we then randomly select
30 injected watermark tuples and use watermark questions (``What is the relationship between $e^{i}_{wm}$ and $e^{j}_{wm}$?'') to query these RAG systems (equipped with various LLMs and knowledge bases) for IP infringement detection. 
The experimental results in Table~\ref{tab:watermark-verification} indicate that RAG-WM achieves effective watermark verification across various LLMs and datasets. The minimum watermark success counts (WSN) for the TREC-COVID, NFCorpus, NQ, HotpotQA, and MS-MARCO datasets are 18, 20, 18, 19, and 18, respectively. These values are far higher than the threshold of 2 required to detect IP infringement of the watermarked RAG. 
Besides, the results also show that the watermark texts are successfully retrieved from the watermarked knowledge database, achieving a watermark information retrieval ratio (WIRR) of greater than or equal to 90\%  across various databases. This high WIRR highlights the effectiveness of our watermark embedding method, which integrates watermark texts into the most relevant database entries to ensure efficient retrieval.  The WIRR for MS-MARCO is lower than for the other knowledge bases, likely due to its large size, which makes watermark retrieval more challenging.


\begin{table}

\begin{threeparttable}
\centering
\footnotesize
\caption{Main Task Performance}
 
\label{tab:main-task-performance}

\begin{tabular}{p{1.3cm}
<{\centering}|p{0.75cm}
<{\centering}|p{1.05cm}
<{\centering}|p{0.7cm}
<{\centering}|p{1.05cm}
<{\centering}|p{0.9cm}
<{\centering}}
\hline
 \multirow{2}{*}{\textbf{Dataset}}
& \textbf{TREC-}& \multirow{2}{*}{\textbf{NFCorpus}} & \multirow{2}{*}{\textbf{NQ}}& \multirow{2}{*}{\textbf{HotpotQA}}& \textbf{MS-} \\

& \textbf{COVID}& & & & \textbf{MARCO} \\ \hline

\textbf{CDPA}& \text{98.00\%}& \text{96.90\%} & \text{99.60\%}& \text{97.70\%}& \text{97.10\% }\\  \hline

\textbf{CIRA}& \text{96.40\%}& \text{89.16\%} & \text{94.66\%}& \text{96.94\%}& \text{98.68\%}\\  \hline

\end{tabular}
\end{threeparttable}

\end{table}


\noindent\textbf{Main Task Performance (Fidelity).} 
We evaluate the clean data performance alignment (i.e., CDPA) between watermarked and clean RAGs using the main task questions from these five datasets. To achieve this, we retrieve relevant texts for these questions from the knowledge bases of the evaluated RAGs and input them into the Llama-2-7B and Vicuna-13B models.  These models are selected for their white-box nature, which allows for easier control (e.g., token sampling strategies) compared to black-box models like GPT-3.5-Turbo and PaLM-2, which are less stable due to factors such as latent variable states\footnote{GPT-3.5-Turbo may produce different answers to the same question because of nondeterministic sampling during inference.}. Moreover, since the evaluation results for Llama-2-7B and Vicuna-13B are similar, we only present the results for Llama-2-7B in Table~\ref{tab:main-task-performance}. We can see that the average CDPA across these tasks is 97.87\%, demonstrating the good performance in maintaining the main task.
Moreover, we also assess the clean information retrieval alignment (i.e., CIRA) between clean and watermarked RAGs. The value is 95.17\% on average. We analyze the primary factor affecting CIRA: when entities in clean queries overlap with those in watermark texts, the watermark texts can interfere with clean information retrieval. However, this interference decreases as the dataset size grows. For example, MS-MARCO, with 8,841,823 texts, has a CIRA of 97.34\%, while NFCorpus, with 3,633 texts, has a CIRA of 89.16\%.

\ignore{
\begin{table}

\begin{threeparttable}
\centering
\footnotesize
\caption{Integrity of RAG-WM \lpz{Put in Appendix or Remove.}}
 
\label{tab:integrity}
\begin{tabular}{m{1cm}
<{\centering}|m{0.9cm}
<{\centering}|m{0.9cm}
<{\centering}|m{0.8cm}
<{\centering}|m{0.8cm}
<{\centering}|m{0.8cm}
<{\centering}}
\hline

\multirow{2}{*}{\textbf{\makecell[c]{Dataset}}} & 
\multirow{2}{*}{\textbf{\makecell[c]{Metrics}}} & 
\multicolumn{4}{c}{\textbf{LLMs}} \\ \cline{3-6} 
 &  & \textbf{GPT-3.5} & \textbf{Llama} & \textbf{Vicuna} & \textbf{PaLM} \\ \hline \hline

\textbf{TREC-COVID}
& \text{WSN}& \text{0} & \text{0}& \text{0}& \text{0} \\ \hline

\textbf{NFCorpus}& \text{WSN}& \text{0} & \text{0}& \text{0}& \text{0} \\  \hline

\textbf{NQ}
& \text{WSN}& \text{0} & \text{0}& \text{0}& \text{0}\\  \hline


\textbf{HotpotQA}
& \text{WSN}& \text{0} & \text{0}& \text{0}& \text{0 }\\  \hline

\textbf{MS-MARCO}
& \text{WSN}& \text{0} & \text{0}& \text{0}& \text{0}\\ \hline

\end{tabular}
\end{threeparttable}

\end{table}
}

\noindent\textbf{Integrity.} We evaluate the integrity of RAG-WM by testing whether it will detect IP infringement over innocent RAGs not stolen from ours. In this experiment, we assess RAG-WM on clean RAGs using four LLMs across five tasks. Ideally, no IP infringement should be detected in these clean RAGs, meaning their WSN should be less than or equal to 2. 
The evaluation results show that the WSN of RAG-WM in clean RAGs is always 0, indicating that RAG-WM does not falsely detect IP infringement in these clean RAGs.




\noindent\textbf{Human Evaluation on Watermark Verification.} The above evaluation results for WSN are obtained with the help of GPT-3.5-Turbo (configured with prompts detailed in Appendix~\ref{sec:prompt-wm-disc}). To validate this approach, we performed a human evaluation. This involves manually verifying whether watermark relationships are contained in the responses produced by the adversary’s deployed LLMs and RAG systems.
The evaluation results are shown in Figure~\ref{fig:human-evalution-wsn} of Appendix. Notably, the results of the LLM-based evaluation align closely with those of the human evaluation, with a difference of no more than 3 in WSN, demonstrating high consistency. Such a high consistency between LLM-based and human evaluations highlights the LLMs as a reliable tool for watermark detection.

\noindent\textbf{Time Consumption.}  The watermarking process introduces additional time compared to a clean RAG. We use GPT-3.5-Turbo to generate the watermark texts. Table~\ref{tab:time} shows that the average time for generating each watermark text is 9.40 seconds, which is acceptable for the owner, as watermarking the RAG's knowledge base is a one-time task for the owner.

\begin{table} [h]
\centering
\footnotesize
\caption{Time Consumption}
 
\label{tab:time}

\begin{tabular}{p{1.3cm}
<{\centering}|p{0.75cm}
<{\centering}|p{1.05cm}
<{\centering}|p{0.5cm}
<{\centering}|p{1.05cm}
<{\centering}|p{0.9cm}
<{\centering}}
\hline
 \multirow{2}{*}{\textbf{Dataset}}
& \textbf{TREC-}& \multirow{2}{*}{\textbf{NFCorpus}} & \multirow{2}{*}{\textbf{NQ}}& \multirow{2}{*}{\textbf{HotpotQA}}& \textbf{MS-} \\

& \textbf{COVID}& & & & \textbf{MARCO} \\ \hline
\textbf{Time }& \multirow{2}{*}{\text{ 5.90}}& \multirow{2}{*}{\text{9.99}} & \multirow{2}{*}{\text{10.11}}&  \multirow{2}{*}{\text{ 10.94}}& \multirow{2}{*}{\text{10.05}}\\  
\textbf{(seconds)} & & & & & \\ \hline

\end{tabular}

\end{table}

\subsection{Impact of Parameters}
\label{subsec:impacts}

The performance of our watermark approach is related to several factors, including the parameters of the RAGs (retriever models, similarity metrics, and the $k$ value for retrieval top $k$ related texts, as defined in equation~(\ref{equ:retrieval})), as well as the parameters of RAG-WM (e.g., the number of injected watermark tuples, the number of watermark texts per tuple, and the watermark queries). We evaluate the impact of these factors using three datasets: REC-COVID, NFCorpus, and MS-MARCO, which differ in scale and knowledge domain. LLaMA-2-7B is used as the adversary's LLM for the RAG system. Due to space limitations, we present the results of impact of watermark queries in Appendix~\ref{subsubsec:watermark-queries}.

\subsubsection{Impact of Retriever Models} 
\begin{figure}[h]
    \centering
    \subfigure[TREC-COVID]{
    \begin{tikzpicture}
    \begin{axis}
        [
        width=3.625cm,
        ymin=-4,
        ymax=38,
        ybar interval=0.4,
        legend style={at={(0.4, 1.0)},anchor=north,draw=none},
        grid=none,
        xtick pos=bottom,
        ytick pos=left,
        symbolic x coords={Contriever,Contriever-ms,ANCE,e}, 
        xtick=data,
        xticklabel style={rotate=-45, anchor=west,font= \tiny,xshift=-5pt},
        ytick={0,10,20,30,38},
        yticklabels={0,10,20,30,\small WSN}, 
        ]
        \addplot[color=blue] coordinates
        {
        (Contriever, 18) (Contriever-ms, 23) (ANCE, 21) (e, 0) 
        };
    \end{axis}
    \end{tikzpicture}
    }
    \subfigure[NFCorpus]{
    \begin{tikzpicture}
    \begin{axis}
        [
         width=3.625cm,
        ymin=-4,
        ymax=38,
        ybar interval=0.4,
        legend style={at={(0.4, 1.0)},anchor=north,draw=none},
        grid=none,
        yticklabels=empty,
        xtick pos=bottom,
        ytick pos=left,
        symbolic x coords={Contriever,Contriever-ms,ANCE,e}, 
        xtick=data,
        xticklabel style={rotate=-45, anchor=west,font= \tiny, xshift=-5pt}
        ]
        \addplot[color=red] coordinates
        {
        (Contriever, 24) (Contriever-ms, 21) (ANCE, 22) (e, 0) 
        };
    \end{axis}
    \end{tikzpicture}
    }
    \subfigure[MS-MARCO]{
    \begin{tikzpicture}
    \begin{axis}
        [
        width=3.625cm,
         ymin=-4,
        ymax=38,
        ybar interval=0.4,
        legend style={at={(0.4, 1.0)},anchor=north,draw=none},
        grid=none,
        yticklabels=empty,
        xtick pos=bottom,
        ytick pos=left,
        symbolic x coords={Contriever,Contriever-ms,ANCE,e}, 
        xtick=data,
        xticklabel style={rotate=-45, anchor=west,font= \tiny,xshift=-5pt}
        ]
        \addplot[color=orange] coordinates
        {
        (Contriever, 19) (Contriever-ms, 18) (ANCE, 15) (e, 0) 
        };
    \end{axis}
    \end{tikzpicture}
    }
    \vspace{-10pt}
    \caption{Impact of Retrievers.    }
    \vspace{-10pt}
    \label{fig:retriever}
\end{figure}
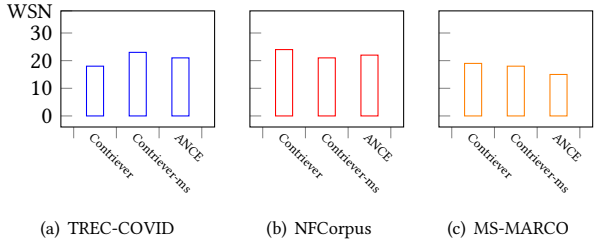

We evaluate three widely used retriever models: Contriever~\cite{izacard2021unsupervised}, Contriever-ms ~\cite{izacard2021unsupervised}, and ANCE ~\cite{xiong2020approximate}. The three retriever models employ mainstream strategies such as unsupervised training, supervised fine-tuning, and retrieval with approximate nearest neighbor optimization, allowing for a thorough performance evaluation.
Figure~\ref{fig:retriever} shows the evaluation results of the effectiveness of RAG-WM on these retrievers. Our results show that RAG-WM consistently performs well across all retrievers, with average watermark success numbers (WSN) of 20, successfully detecting IP infringement. 
This is because the watermark texts are semantically aligned with the watermark queries, achieving an average retrieval rate of 95.93\% (i.e., WIRR), making them more likely to be retrieved by various retrievers.

\subsubsection{Impact of Similarity Metrics} 
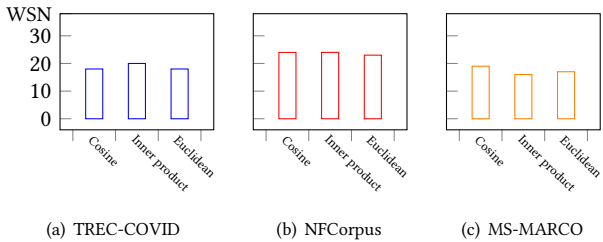
\begin{figure}[h]
    \centering
    \subfigure[TREC-COVID]{
    \begin{tikzpicture}
    \begin{axis}
        [
        width=3.625cm,
        ymin=-4,
        ymax=38,
        ybar interval=0.4,
        legend style={at={(0.4, 1.0)},anchor=north,draw=none},
        grid=none,
        xtick pos=bottom,
        ytick pos=left,
        symbolic x coords={Cosine,Inner product,Euclidean,e}, 
        xtick=data,
        xticklabel style={rotate=-45, anchor=west,font= \tiny,xshift=-5pt},
        ytick={0,10,20,30,38},
        yticklabels={0,10,20,30,\small WSN}, 
        ]
        \addplot[color=blue] coordinates
        {
        (Cosine, 18) (Inner product, 20) (Euclidean, 18) (e, 0) 
        };
    \end{axis}
    \end{tikzpicture}
    }
            \hspace{-0.2cm}
    \subfigure[NFCorpus]{
    \begin{tikzpicture}
    \begin{axis}
        [
        width=3.625cm,
        ymin=-4,
        ymax=38,
        ybar interval=0.4,
        legend style={at={(0.4, 1.0)},anchor=north,draw=none},
        grid=none,
        yticklabels=empty,
        xtick pos=bottom,
        ytick pos=left,
        symbolic x coords={Cosine,Inner product,Euclidean,e}, 
        xtick=data,
        xticklabel style={rotate=-45, anchor=west,font= \tiny, xshift=-5pt}
        ]
        \addplot[color=red] coordinates
        {
        (Cosine, 24) (Inner product, 24) (Euclidean, 23) (e, 0) 
        };
    \end{axis}
    \end{tikzpicture}
    }
        \hspace{-0.2cm}
    \subfigure[MS-MARCO]{
    \begin{tikzpicture}
    \begin{axis}
        [
        width=3.625cm,
        ymin=-4,
        ymax=38,
        ybar interval=0.4,
        legend style={at={(0.4, 1.0)},anchor=north,draw=none},
        grid=none,
        yticklabels=empty,
        xtick pos=bottom,
        ytick pos=left,
        symbolic x coords={Cosine,Inner product,Euclidean,e}, 
        xtick=data,
        xticklabel style={rotate=-45, anchor=west,font= \tiny,xshift=-5pt}
        ]
        \addplot[color=orange] coordinates
        {
        (Cosine, 19) (Inner product, 16) (Euclidean, 17) (e, 0) 
        };
    \end{axis}
    \end{tikzpicture}
    }
    \vspace{-10pt}
    \caption{Impact of Similarity Metrics.    }
    \vspace{-10pt}
    \label{fig:similarity-metrics}
\end{figure}

To assess the impact of similarity metrics, we evaluate three metrics: cosine similarity, inner product, and Euclidean distance to calculate the similarity of embedding vectors when retrieving texts from a watermarked knowledge database for a watermark query. 
Figure~\ref{fig:similarity-metrics} shows the results of these similarity metrics. We observe that RAG-WM produces consistent results across the three tasks. Specifically, the WSN values are similar for each similarity metric across tasks, with differences of no more than 3 for the same task. Furthermore, these WSN values are significantly higher than 2, effectively protecting the IP of RAGs.

\subsubsection{Impact of k}
\label{subsubsec:impact-k}
\begin{figure}[h]
    \centering
    \subfigure[TREC-COVID]{
    \begin{tikzpicture}
    \begin{axis}[
        width=3.625cm,
        axis y line=left,
        ylabel style = {name=ylabel1}, ymax=34,ymin=11,
        legend style={nodes={scale=0.7, transform shape},at={(0.99,0.20)},anchor=east},
        ytick={15,20,25,30,34},
        yticklabels={15, 20, 25,30, WSN}, 
        xtick={1,2,3,4,5}, 
        xticklabel style={font= \small},
        xlabel={$k$ Value}, 
        xlabel style={font=\small,opacity=0},
        yticklabel style={font=\small},
    ]

        \addplot[mark=o,every mark/.append style={solid},color=blue]
        plot coordinates { 
            (1,18)
            (2,23)
            (3,25)
            (4,24)
            (5,24)
        };
        
    \end{axis}
    
    \begin{axis}[
        width=3.625cm,
        axis y line=right,
        axis x line=none, ylabel style = {name=ylabel2}, ymax=103,ymin=87,
        yticklabels={}, 
        ytick style={draw=none},
    ]

        \addplot[dashed,mark=o,every mark/.append style={solid},color=red]
        plot coordinates { 
            (1,96.67)
            (2,100.00)
            (3,100.00)
            (4,100.00)
            (5,100.00)
        };
    \end{axis}
    \end{tikzpicture}
    }
    \hspace{-0.8cm}
    \subfigure[NFCorpus]{
    \begin{tikzpicture}
       \begin{axis}[
        width=3.625cm,
        axis y line=left, 
        ylabel style = {name=ylabel1}, ymax=34,ymin=11,
        legend style={nodes={scale=0.5, transform shape},at={(0.80,0.20)},anchor=east},
        xtick={1,2,3,4,5}, 
        yticklabels={}, 
        ytick style={draw=none},
        xticklabel style={font= \small},
        xlabel={$k$ Value}, 
        xlabel style={font=\small},
    ]
        \addlegendimage{red,dashed,mark=o}
        \addlegendentry{WIRR}
        \addlegendimage{blue,solid,mark=o}
        \addlegendentry{WSN}
        
        \addplot[mark=o,every mark/.append style={solid},color=blue]
        plot coordinates { 
            (1,24)
            (2,26)
            (3,27)
            (4,24)
            (5,27)
        };

    \end{axis}
    
    \begin{axis}[
        width=3.625cm,
        axis y line=right,
        axis x line=none, ylabel style = {name=ylabel2}, ymax=103,ymin=87,
        yticklabels={}, 
        ytick style={draw=none},
    ]

        \addplot[dashed,mark=o,every mark/.append style={solid},color=red]
        plot coordinates { 
            (1,100.00)
            (2,100.00)
            (3,100.00)
            (4,100.00)
            (5,100.00)
        };

    \end{axis}
    \end{tikzpicture}
    }
    \hspace{-0.8cm}
    \subfigure[MS-MARCO]{
    \begin{tikzpicture}
    \begin{axis}[
        width=3.625cm,
        axis y line=left,
        ylabel style = {name=ylabel1}, ymax=34,ymin=11,
        legend style={nodes={scale=0.5, transform shape},at={(0.99,0.19)},anchor=east},
        xtick={1,2,3,4,5}, 
        yticklabels={}, 
        ytick style={draw=none},
        xticklabel style={font= \small},
        xlabel={$k$ Value}, 
        xlabel style={font=\small,opacity=0},
    ]

        \addplot[mark=o,every mark/.append style={solid},color=blue]
        plot coordinates { 
            (1,19)
            (2,23)
            (3,22)
            (4,21)
            (5,23)
        };

    \end{axis}
    
    \begin{axis}[
        width=3.625cm,
        axis y line=right,
        axis x line=none, ylabel style = {name=ylabel2}, ymax=103,ymin=87,
        ytick={90, 95,100,103},
        yticklabels={90\%, 95\%, 100\%,WIRR}, 
        legend style={draw=none, fill=none, text=none},
        yticklabel style={font=\small},
    ]

        \addplot[dashed,mark=o,every mark/.append style={solid},color=red]
        plot coordinates { 
            (1,96.67)
            (2,96.67)
            (3,100.00)
            (4,100.00)
            (5,100.00)
        };

    \end{axis}
    \end{tikzpicture}
    }
\vspace{-10pt}
    \caption{Impact of $k$.
    }
    \vspace{-10pt}
    \label{fig:impact-of-k}
\end{figure}
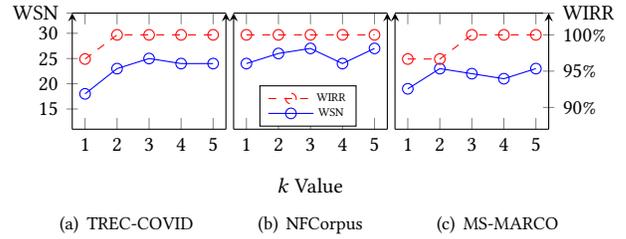

RAG returns the top $k$ text records most relevant to the querying question to LLMs. This parameter is typically set by the adversary. We evaluate the impact of $k$ by setting its value within the range of $[1, 5]$. The evaluation results are shown in Figure~\ref{fig:impact-of-k}.
We can see that the WSN of RAG-WM remains significantly high (well above the threshold of 2) when $k$ is between $1$ and $5$. This is because most of the retrieved texts contain watermark information, resulting in a high WIRR, 99.11\% on average.  
Additionally, with an increase in the number of retrieved texts ($k$), both the WSN and WIRR increase. For example, the WSN and WIRR are typically higher for $k=5$ than for $k=1$. Thus, we use $k=1$ in other evaluations, as it represents the worst-case scenario.

\subsubsection{The Number of Injected Watermark Tuples}
Watermark tuples contain the watermark information, i.e., the embedded entities and their relationships. The number of injected watermark tuples can affect the subsequent watermark verification process. To detect IP infringement, the owner queries a suspicious RAG system with 30 randomly selected watermark questions related to the injected watermark tuples. We assess how the number of embedded watermark tuples, specifically 40, 50, 60, 80, and 100, affects verification performance. The evaluation results are in Figure~\ref{fig:watermark-tuples}.
The verification success rates remain stable across the 30 queries, with the WSN consistently around 20 and the WIRR exceeding 95\%, regardless of the number of embedded watermark tuples (40, 50, 60, 80, or 100). These results suggest that our watermark achieves stable performance with a limited number of tuples. 
The main task performance (i.e., CDPA) remains stable, exceeding 94\% when embedding different numbers of watermark tuples. As the number of watermark tuples increases, the clean information retrieval alignment (CIRA) decreases in the NFCorpus task, due to the injection of more watermark texts and the smaller size of its knowledge base. In contrast, for larger knowledge bases (e.g., TREC-COVID and MS-MARCO), the CIRA remains stable because the number of watermark texts is minimal compared to the total number of task-related texts.
Based on these evaluations, we default to embedding 50 tuples in the experiments of this paper, as this number achieves good performance.

\begin{figure}[h]
    \centering    
    \subfigure[TREC-COVID]{

    \begin{tikzpicture}
    \begin{axis}[
             width=3.625cm,
        axis y line=left,
        xlabel={$\text{Number}$}, 
        xlabel style={font=\small, opacity=0},
        ylabel style = {name=ylabel1,yshift=-0.5cm}, ymax=34,ymin=11,
        legend style={nodes={scale=0.7, transform shape},at={(0.99,0.20)},anchor=east},
        xtick={40,60,80,100}, 
        ytick={15,20,25,30,34},
        yticklabels={15, 20, 25,30, WSN}, 
        yticklabel style={font=\small},
        xticklabel style={font= \small},
    ]

        \addplot[mark=o,every mark/.append style={solid},color=blue]
        plot coordinates { 
            (40,21)
            (50,18)
            (60,22)
            (80,22)
            (100,23)
        };
        
    \end{axis}
    
    \begin{axis}[
            width=3.625cm,
        axis y line=right,
        axis x line=none, ylabel style = {name=ylabel2}, ymax=103,ymin=87,
        yticklabels={}, 
        ytick style={draw=none}
    ]

        \addplot[dashed,mark=o,every mark/.append style={solid},color=red]
        plot coordinates { 
            (40,100.00)
            (50,96.67)
            (60,100.00)
            (80,100.00)
            (100,96.67)
        };
    \end{axis}
    \end{tikzpicture}
    }
    \hspace{-0.8cm}
    \subfigure[NFCorpus]{
    \begin{tikzpicture}
       \begin{axis}[
        width=3.625cm,
        axis y line=left, 
        xlabel={$\text{Number}$}, 
      xlabel style={font=\small},
        ylabel style = {name=ylabel1},ymax=34,ymin=11,
        legend style={nodes={scale=0.5, transform shape},at={(0.80,0.20)},anchor=east},
        xtick={40,60,80,100},
        yticklabels={}, 
        ytick style={draw=none},
        xticklabel style={font= \small}
    ]
        \addlegendimage{red,dashed,mark=o}
        \addlegendentry{WIRR}
        \addlegendimage{blue,solid,mark=o}
        \addlegendentry{WSN}

        \addplot[mark=o,every mark/.append style={solid},color=blue]
        plot coordinates { 
            (40,19)
            (50,24)
            (60,21)
            (80,23)
            (100,19)
        };

    \end{axis}
    
    \begin{axis}[
            width=3.625cm,
        axis y line=right,
        axis x line=none, ylabel style = {name=ylabel2}, ymax=103,ymin=87,
        yticklabels={}, 
        ytick style={draw=none},
    ]

        \addplot[dashed,mark=o,every mark/.append style={solid},color=red]
        plot coordinates { 
            (40,100.00)
            (50,100.00)
            (60,100.00)
            (80,100.00)
            (100,100.00)
        };

    \end{axis}
    \end{tikzpicture}
    }
    \hspace{-0.8cm}
    \subfigure[MS-MARCO]{
    \begin{tikzpicture}
    \begin{axis}[
            width=3.625cm,
        axis y line=left,
        xlabel={$\text{Number}$}, 
        xlabel style={font=\small,opacity=0},
        ylabel style = {name=ylabel1}, ymax=34,ymin=11,
        legend style={nodes={scale=0.7, transform shape},at={(0.99,0.51)},anchor=east},
        xtick={40,60,80,100},
        yticklabels={}, 
        ytick style={draw=none},
        xticklabel style={font= \small},
    ]

        \addplot[mark=o,every mark/.append style={solid},color=blue]
        plot coordinates { 
            (40,19)
            (50,19)
            (60,20)
            (80,18)
            (100,18)
        };

    \end{axis}
    
    \begin{axis}[
            width=3.625cm,
        axis y line=right,
        axis x line=none, ylabel style = {name=ylabel2m,yshift=0.2cm}, ymax=103,ymin=87,
        ytick={90, 95,100,103},
        yticklabels={90\%, 95\%, 100\%, WIRR}, 
        yticklabel style={font=\small},
    ]

        \addplot[dashed,mark=o,every mark/.append style={solid},color=red]
        plot coordinates { 
            (40,100.00)
            (50,96.67)
            (60,96.67)
            (80,96.67)
            (100,100.00)
        };

    \end{axis}
    \end{tikzpicture}
    }


    
    \subfigure[TREC-COVID]{
    \begin{tikzpicture}
    \begin{axis}[
                  width=3.625cm,
        axis y line=left,
        ylabel style = {name=ylabel1,yshift=-0.2cm}, ymax=105,ymin=80,
        legend style={nodes={scale=0.7, transform shape},at={(0.99,0.20)},anchor=east},
        xtick={40,60,80,100},
        ytick={85,90, 95,100,105},
        yticklabels={85\%,90\%, 95\%, 100\%,CIRA}, 
        xticklabel style={font= \small},
        yticklabel style={font=\small},
        xlabel={$\text{Number}$}, 
        xlabel style={font=\small,opacity=0},
    ]

        \addplot[mark=triangle,every mark/.append style={solid},color=red]
        plot coordinates { 
            (40,95.20)
            (50,96.40)
            (60,94.40)
            (80,94.40)
            (100,93.60)
        };
        
    \end{axis}
    
    \begin{axis}[
             width=3.625cm,
        axis y line=right,
        axis x line=none, ylabel style = {name=ylabel2}, ymax=103,ymin=87,
        yticklabels={}, 
        ytick style={draw=none},
    ]

        \addplot[mark=triangle,every mark/.append style={solid},color=blue]
        plot coordinates { 
            (40,98.00)
            (50,98.00)
            (60,96.00)
            (80,98.00)
            (100,98.00)
        };
    \end{axis}
    \end{tikzpicture}
    }
    \hspace{-0.8cm}
    \subfigure[NFCorpus]{
    \begin{tikzpicture}
    \begin{axis}[
             width=3.625cm,
        axis y line=left,
        ylabel style = {name=ylabel1,font= \small},  ymax=105,ymin=80,
        legend style={nodes={scale=0.5, transform shape},at={(0.80,0.82)},anchor=east},
        xtick={40,60,80,100}, 
        yticklabels={}, 
        ytick style={draw=none},
        xticklabel style={font= \small},
        xlabel={$\text{Number}$}, 
        xlabel style={font=\small},
    ]
        \addlegendimage{blue,solid,mark=triangle}
        \addlegendentry{CDPA}
        \addlegendimage{red,solid,mark=triangle}
        \addlegendentry{CIRA}
        
        \addplot[mark=triangle,every mark/.append style={solid},color=red]
        plot coordinates { 
            (40,91.27)
            (50,89.16)
            (60,87.28)
            (80,86.01)
            (100,85.30)
        };

    \end{axis}
    
    \begin{axis}[
               width=3.625cm,
        axis y line=right,
        axis x line=none, ylabel style = {name=ylabel2}, ymax=103,ymin=87,
          yticklabels={}, 
        ytick style={draw=none},
    ]

        \addplot[mark=triangle,every mark/.append style={solid},color=blue]
        plot coordinates { 
            (40,94.74)
            (50,96.90)
            (60,94.43)
            (80,96.91)
            (100,96.60)
        };

    \end{axis}
    \end{tikzpicture}
    }
    \hspace{-0.8cm}
    \subfigure[MS-MARCO]{
    \begin{tikzpicture}
    \begin{axis}[
             width=3.625cm,
        axis y line=left,
        ylabel style = {name=ylabel1},  ymax=105,ymin=80,
        legend style={nodes={scale=0.5, transform shape},at={(0.99,0.19)},anchor=east},
        xtick={40,60,80,100}, 
        yticklabels={}, 
        ytick style={draw=none},
        xticklabel style={font= \small},
                xlabel={$\text{Number}$}, 
        xlabel style={font=\small,opacity=0},
    ]
        
        \addplot[mark=triangle,every mark/.append style={solid},color=red]
        plot coordinates { 
            (40,96.10)
            (50,98.68)
            (60,96.78)
            (80,95.24)
            (100,96.20)
        };

    \end{axis}

    \begin{axis}[
                 width=3.625cm,
        axis y line=right,
        axis x line=none, ylabel style = {name=ylabel2,yshift=0.2cm,font= \small}, ymax=103,ymin=87,
        ytick={90, 95,100,103},
        yticklabels={90\%, 95\%, 100\%,CDPA}, 
                yticklabel style={font=\small},
        legend style={draw=none, fill=none, text=none}
    ]

        \addplot[mark=triangle,every mark/.append style={solid},color=blue]
        plot coordinates { 
            (40,98.00)
            (50,97.10)
            (60,98.40)
            (80,97.10)
            (100,97.70)
        };

    \end{axis}
    \end{tikzpicture}
    }
    \vspace{-10pt}
    \caption{Impact of Number of Watermark Tuples.}
    \vspace{-10pt}
    \label{fig:watermark-tuples}
\end{figure}
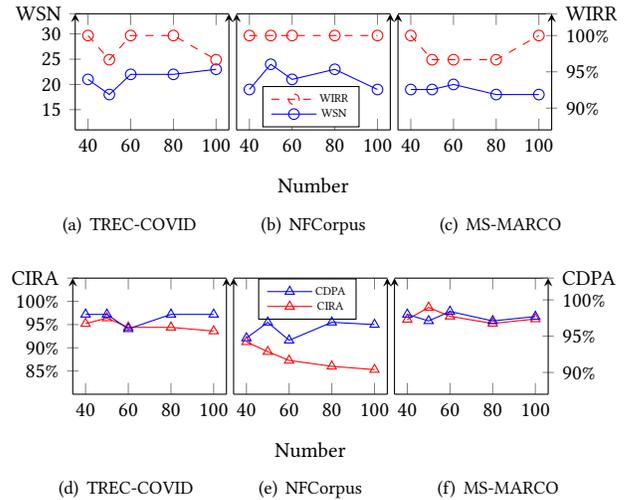

\begin{figure}[h]
    \centering
    \subfigure[TREC-COVID]{
    \begin{tikzpicture}
    \begin{axis}[
               width=3.625cm,
        axis y line=left,
        ylabel style = {name=ylabel1},ymax=34,ymin=11,
        legend style={nodes={scale=0.5, transform shape},at={(0.65,0.20)},anchor=east},
        xtick={1,2,3,4,5}, 
        ytick={15,20,25,30,34},
        yticklabels={15, 20, 25,30, WSN}, 
        xlabel={$\text{Number}$}, 
        xlabel style={font=\small,opacity=0},
        xticklabel style={font= \small},
        yticklabel style={font=\small},
    ]
    
        \addplot[mark=o,every mark/.append style={solid},color=blue]
        plot coordinates { 
            (1,16)
            (2,16)
            (3,16)
            (4,19)
            (5,19)
        };
        
    \end{axis}
    
    \begin{axis}[
               width=3.625cm,
        axis y line=right,
        axis x line=none, ylabel style = {name=ylabel2},  ymax=103,ymin=87,
        yticklabels={}, 
        ytick style={draw=none},
        yticklabel style={font=\small},
    ]

        \addplot[mark=triangle,every mark/.append style={solid},color=red]
        plot coordinates { 
            (1,100.00)
            (2,98.00)
            (3,98.00)
            (4,100.00)
            (5,98.00)
        };
    \end{axis}
    \end{tikzpicture}
    }
    \hspace{-0.8cm}
    \subfigure[NFCorpus]{
    \begin{tikzpicture}
       \begin{axis}[
                  width=3.625cm,
        axis y line=left, 
        ylabel style = {name=ylabel1}, ymax=34,ymin=11,
        legend style={nodes={scale=0.5, transform shape},at={(0.99,0.80)},anchor=east},
        xtick={1,2,3,4,5}, 
        yticklabels={}, 
        ytick style={draw=none},
        xlabel={$\text{Number}$}, 
        xlabel style={font=\small},
        xticklabel style={font= \small},
    ]
        \addlegendimage{blue,mark=o}
        \addlegendentry{WSN}
        \addlegendimage{red,solid,mark=triangle}
        \addlegendentry{CDPA}
        
        \addplot[mark=o,every mark/.append style={solid},color=blue]
        plot coordinates { 
            (1,18)
            (2,18)
            (3,19)
            (4,19)
            (5,21)
        };

    \end{axis}
    
    \begin{axis}[
               width=3.625cm,
        axis y line=right,
        axis x line=none, ylabel style = {name=ylabel2}, ymax=103,ymin=87,
        yticklabels={}, 
        ytick style={draw=none},
    ]

        \addplot[mark=triangle,every mark/.append style={solid},color=red]
        plot coordinates { 
            (1,95.98)
            (2,95.67)
            (3,95.98)
            (4,95.35)
            (5,95.98)
        };

    \end{axis}
    \end{tikzpicture}
    }
    \hspace{-0.8cm}
    \subfigure[MS-MARCO]{
    \begin{tikzpicture}
    \begin{axis}[
               width=3.625cm,
        axis y line=left,
        ylabel style = {name=ylabel1}, ymax=34,ymin=11,
        legend style={nodes={scale=0.5, transform shape},at={(0.99,0.19)},anchor=east},
        xtick={1,2,3,4,5}, 
        yticklabels={}, 
        ytick style={draw=none},
                xlabel={$\text{Number}$}, 
        xlabel style={font=\small,opacity=0},
                xticklabel style={font= \small},
    ]

        \addplot[mark=o,every mark/.append style={solid},color=blue]
        plot coordinates { 
            (1,14)
            (2,14)
            (3,15)
            (4,19)
            (5,19)
        };

    \end{axis}
    
    \begin{axis}[
                  width=3.625cm,
        axis y line=right,
        axis x line=none, ylabel style = {name=ylabel2}, ymax=103,ymin=87,
        ytick={90, 95,100,103},
        yticklabels={90\%, 95\%, 100\%,CDPA}, 
        legend style={draw=none, fill=none, text=none},
        yticklabel style={font=\small},
    ]

        \addplot[mark=triangle,every mark/.append style={solid},color=red]
        plot coordinates { 
            (1,98.70)
            (2,98.20)
            (3,98.50)
            (4,98.20)
            (5,98.10)
        };

    \end{axis}
    \end{tikzpicture}
    }
    \caption{Impact of Number of Injected Texts per Watermark
Tuple.}
    \vspace{-10pt}
    \label{fig:the-number-of-injected-texts}
\end{figure}
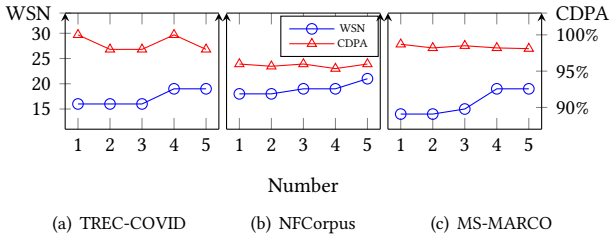

\subsubsection{The Number of Injected Texts per Watermark Tuple} For a given question, the RAG system retrieves the top $k$ most relevant texts. For each watermark tuple, we can generate multiple texts with the same semantics but different content, thereby increasing the proportion of watermark texts retrieved.
We evaluate the impact of varying the number (i.e., $N_{wm}$) of injected watermark texts per tuple. Specifically, we set $k=1$ (i.e., the number of retrieved most related texts) and assess watermark performance for $N_{wm}$ values ranging from 1 to 5.  Figure~\ref{fig:the-number-of-injected-texts} shows the results.
We observe that the WSN value increases as the number of injected watermark texts increases. This is because adding more watermark texts raises the likelihood of retrieving the watermark text, making the watermark relation more likely to appear in the LLM outputs. Therefore, we can generate multiple watermark texts for each watermark tuple and inject them into RAG to improve watermark performance. Moreover, adversaries cannot reliably detect multiple watermark texts for the same watermark tuple using duplicate text filtering, as demonstrated in Section~\ref{subsubsec:duplicate}).
Additionally, CDPA remains stable for $ N_{wm} =[1,5]$, consistently above 95\%, indicating good performance in maintaining the main task.

\subsection{Robustness}
\label{subsec:robustness}

After stealing RAGs, attackers may attempt to remove the watermark. The adversary can utilize paraphrasing, removing unrelated content, inserting knowledge, and expanding knowledge attacks.

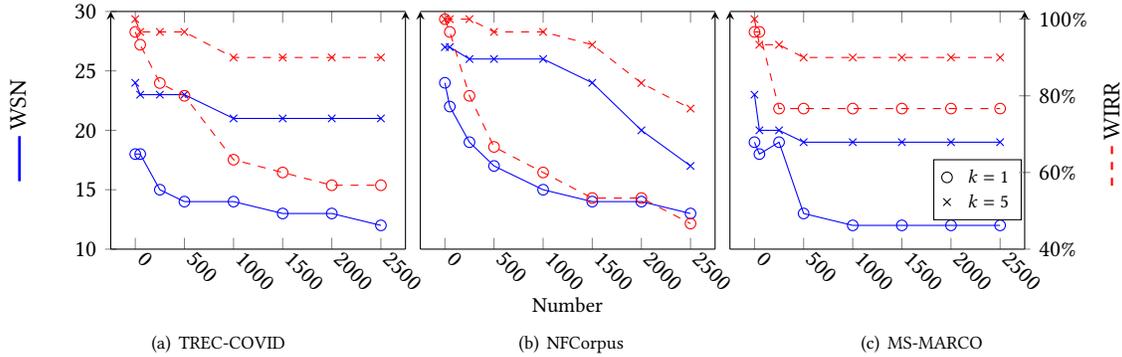
\begin{figure*}[t]
    \centering
    \subfigure[TREC-COVID]{
    \begin{tikzpicture}
    \begin{axis}[
        width=5.5cm,
        axis y line=left,
        ylabel={{\color{blue}\hdashrule[0.5ex]{0.6cm}{1pt}{3mm 0pt}} $\text{WSN}$}, 
        ylabel style = {name=ylabel1}, ymax=30,ymin=10,
        legend style={nodes={scale=0.7, transform shape},at={(0.99,0.20)},anchor=east},
        xtick={0,500,1000,1500,2000,2500}, 
        xticklabels={0,500,1000,1500,2000,2500} ,
        xticklabel style={rotate=-45, anchor=west},
        xlabel={$\text{Number}$}, 
        xlabel style={font=\small,opacity=0},
    ]

        \addplot[mark=o,every mark/.append style={solid},color=blue]
        plot coordinates { 
            (0,18)
            (50,18)
            (250,15)
            (500,14)
            (1000,14)
            (1500,13)
            (2000,13)
            (2500,12)
        };
        \addplot[mark=x,every mark/.append style={solid},color=blue]
        plot coordinates { 
            (0,24)
            (50,23)
            (250,23)
            (500,23)
            (1000,21)
            (1500,21)
            (2000,21)
            (2500,21)
        };
        
    \end{axis}
    
    \begin{axis}[
        width=5.5cm,
        axis y line=right,
        axis x line=none, ylabel style = {name=ylabel2}, ymax=102,ymin=40,
        yticklabels={}, 
        ytick style={draw=none},
    ]

        \addplot[dashed,mark=o,every mark/.append style={solid},color=red]
        plot coordinates { 
            (0,96.67)
            (50,93.33)
            (250,83.33)
            (500,80.00)
            (1000,63.33)
            (1500,60.00)
            (2000,56.67)
            (2500,56.67)
        };

        \addplot[dashed,mark=x,every mark/.append style={solid},color=red]
        plot coordinates { 
            (0,100.00)
            (50,96.67)
            (250,96.67)
            (500,96.67)
            (1000,90.00)
            (1500,90.00)
            (2000,90.00)
            (2500,90.00)
        };
    \end{axis}
    \end{tikzpicture}
    }
    \hspace{-0.8cm}
    \subfigure[NFCorpus]{
    \begin{tikzpicture}
       \begin{axis}[
        width=5.5cm,
        axis y line=left, 
        ylabel style = {name=ylabel1}, ymax=30,ymin=10,
        legend style={nodes={scale=0.7, transform shape},at={(0.99,0.20)},anchor=east},
        xtick={0,500,1000,1500,2000,2500}, 
        xticklabels={0,500,1000,1500,2000,2500},
        yticklabels={}, 
        xticklabel style={rotate=-45, anchor=west},
        ytick style={draw=none},
        xlabel={$\text{Number}$}, 
        xlabel style={font=\small},
    ]
        
        \addplot[mark=o,every mark/.append style={solid},color=blue]
        plot coordinates { 
            (0,24)
            (50,22)
            (250,19)
            (500,17)
            (1000,15)
            (1500,14)
            (2000,14)
            (2500,13)
        };
        \addplot[mark=x,every mark/.append style={solid},color=blue]
        plot coordinates { 
            (0,27)
            (50,27)
            (250,26)
            (500,26)
            (1000,26)
            (1500,24)
            (2000,20)
            (2500,17)
        };

    \end{axis}
    
    \begin{axis}[
         width=5.5cm,
        axis y line=right,
        axis x line=none, ylabel style = {name=ylabel2}, ymax=102,ymin=40,
        yticklabels={}, 
        ytick style={draw=none},
    ]

        \addplot[dashed,mark=o,every mark/.append style={solid},color=red]
        plot coordinates { 
            (0,100.00)
            (50,96.67)
            (250,80.00)
            (500,66.67)
            (1000,60.00)
            (1500,53.33)
            (2000,53.33)
            (2500,46.67)
        };
        \addplot[dashed,mark=x,every mark/.append style={solid},color=red]
        plot coordinates { 
            (0,100.00)
            (50,100.00)
            (250,100.00)
            (500,96.67)
            (1000,96.67)
            (1500,93.33)
            (2000,83.33)
            (2500,76.67)
        };
    \end{axis}
    \end{tikzpicture}
    }
    \hspace{-0.8cm}
    \subfigure[MS-MARCO]{
    \begin{tikzpicture}
    \begin{axis}[
         width=5.5cm,
        axis y line=left,
        ylabel style = {name=ylabel1}, ymax=30,ymin=10,
        legend style={nodes={scale=0.8, transform shape},at={(0.99,0.25)},anchor=east, column sep=0.5em},
        xtick={0,500,1000,1500,2000,2500}, 
        xticklabels={0,500,1000,1500,2000,2500},
        yticklabels={}, 
        ytick style={draw=none},
        xticklabel style={rotate=-45, anchor=west},
        xlabel = {$\text{Number}$}, 
        xlabel style={font=\small,opacity=0},
    ]

        \addlegendimage{mark=o,only marks}
        \addlegendentry{$k=1$}
        \addlegendimage{mark=x,only marks}
        \addlegendentry{$k=5$}
        
        \addplot[mark=o,every mark/.append style={solid},color=blue]
        plot coordinates { 
            (0,19)
            (50,18)
            (250,19)
            (500,13)
            (1000,12)
            (1500,12)
            (2000,12)
            (2500,12)
        };
        \addplot[mark=x,every mark/.append style={solid},color=blue]
        plot coordinates { 
            (0,23)
            (50,20)
            (250,20)
            (500,19)
            (1000,19)
            (1500,19)
            (2000,19)
            (2500,19)
        };

    \end{axis}
    
    \begin{axis}[
        width=5.5cm,
        ylabel={\begin{tikzpicture}
            \draw[red, thick, dashed] (0,0) -- (0.6cm,0);
            \node at (1.0cm,0) {$\text{WIRR}$};
        \end{tikzpicture}},
        axis y line=right,
        axis x line=none, ylabel style = {name=ylabel2}, ymax=102,ymin=40,
        ytick={40,60,80,100},
        yticklabels={40\%,60\%,80\%,100\%}, 
        legend style={ fill=none, text=none},
    ]

        \addplot[dashed,mark=o,every mark/.append style={solid},color=red]
        plot coordinates { 
            (0,96.67)
            (50,96.67)
            (250,76.67)
            (500,76.67)
            (1000,76.67)
            (1500,76.67)
            (2000,76.67)
            (2500,76.67)
        };
        \addplot[dashed,mark=x,every mark/.append style={solid},color=red]
        plot coordinates { 
            (0,100.00)
            (50,93.33)
            (250,93.33)
            (500,90.00)
            (1000,90.00)
            (1500,90.00)
            (2000,90.00)
            (2500,90.00)
        };
    \end{axis}
    \end{tikzpicture}
    }
    \vspace{-5pt}
    \caption{Knowledge Insertion Attack.}
    \vspace{-5pt}
    \label{fig:knowledge-insertion-attack}
\end{figure*}

\subsubsection{Paraphrasing Attack}
\label{subsec:paraphrasing-attack}

Paraphrasing has been employed as a strategy to evade watermark detection in LLM-generated watermark texts~\cite{krishna2024paraphrasing, kirchenbauer2023reliability, christ2024undetectable, kirchenbauer2023reliability}, thus it may be an effective technique to evade the watermark verification in RAG-WM. 
Specifically, when a watermark query is issued, the adversary can use an LLM to automatically paraphrase the retrieved texts from the watermarked knowledge database, thereby removing the watermark information and further evading verification. For paraphrasing, 
we follow the approach in~\cite{kirchenbauer2023reliability}, 
employing GPT-3.5-Turbo with the prompt ``paraphrase the following sentences'', a temperature setting of 0.7, and a maximum output length of 200 tokens.
Table~\ref{tab:paraphrase}  shows the results of the attack. The average WSN value is 14, well above 2, indicating that RAG-WM remains detectable even after paraphrasing. This suggests that paraphrasing cannot effectively remove our RAG-WM. 
This is because paraphrasing can modify the wording and structure of the text, but it cannot alter the knowledge (i.e., the entities and relationship types) present in the content. In contrast,~\cite{jovanovic2024ward}, which is based on a red-green list watermark, is not robust against paraphrasing attacks. The adversary can paraphrase any word in the RAG texts by replacing words from the green list with those from the red list, rendering the WARD watermark ineffective.

\begin{table}[ht]
\centering
\begin{threeparttable}

\footnotesize
\caption{Paraphrasing Attack}
 
\label{tab:paraphrase}

\begin{tabular}{m{1cm}
<{\centering}|m{2cm}
<{\centering}|m{1cm}
<{\centering}|m{1.5cm}
<{\centering}}
\hline

\textbf{Dataset} & \textbf{TREC-COVID} & \textbf{NFCorpus} & \textbf{MS-MARCO}   \\ \hline \hline

\textbf{WSN} & \text{13}& \text{17}& \text{12}\\ \hline


\end{tabular}
\end{threeparttable}

\end{table}

\subsubsection{Unrelated Content Removal}

Watermark content serves as extra information for detecting IP infringement. Such content may be removed by excluding specific sentences from the retrieved content in RAGs. Since removing content closely related to the main task would degrade performance, an adversary would desire to remove only unrelated content to minimize the impact on the main task. To attack RAG-WM, we adapt this method to remove such sentences from retrieved texts from the watermarked knowledge database for a watermark query. Specifically, GPT-3.5-Turbo is used to remove unrelated content based on a designed prompt (details in Appendix~\ref{sec:unrelated}). The experimental results are shown in Table~\ref{tab:unrelated-content-removal}.
After the attack, the average WSN is 12, satisfying IP infringement detection. Thus, RAG-WM is robust against unrelated content removal.

\begin{table}[ht]
\centering
\begin{threeparttable}

\footnotesize
\caption{Unrelated Content Removal Attack}
 
\label{tab:unrelated-content-removal}

\begin{tabular}{m{1cm}
<{\centering}|m{2cm}
<{\centering}|m{1cm}
<{\centering}|m{1.5cm}
<{\centering}}
\hline

\textbf{Dataset} & \textbf{TREC-COVID} & \textbf{NFCorpus} & \textbf{MS-MARCO}   \\ \hline \hline

\textbf{WSN} & \text{12}& \text{14}& \text{10}\\ \hline


\end{tabular}
\end{threeparttable}

\end{table}

 







\subsubsection{Knowledge Insertion Attack }
Adversaries can insert misleading information (acting as noise) into an RAG's knowledge base to interfere with the watermark retrieval process~\cite{cuconasu2024power}. Since our watermark text encode information about two entities and their relation, the adversary can maximize the interference by inserting records with randomly selected entities and relations. Such insertion may involve the watermark entities and introduce random relations (different from the watermark relations), potentially causing the failure of watermark relation retrieval.
To launch such an attack, we generate attack texts by randomly combining two texts from the watermarked RAG knowledge base and directly injecting them into the watermarked RAG. 
We vary the number of inserted texts from 0 to 2500 under default settings and retrieve the top relevant texts (evaluating both top~1 and top~5 retrieval cases) for each watermark query. The results are shown in Figure~\ref{fig:knowledge-insertion-attack}.
In the top~1 case, with no texts injected, the WSN values for TREC-COVID, NFCorpus, and MS-MARCO are 18, 24, and 19, respectively. With 2,500 injected texts, the WSN values decrease to 12, 13, and 12.
While the WSN decreases as the number of injected texts increases, it remains above 2, indicating that the attack does not compromise the IP infringement detection of RAG-WM. The decrease in WSN is mainly due to the interference from the injected texts, which affects the retrieval of watermark texts. For instance, when 2,500 texts are injected (approximately 70\% of the original NFCorpus texts), the WIRR gradually drops from 100.00\% to 46.67\%.
Furthermore, we observe that the decrease in WSN is smaller when retrieval is performed using the top~5 texts compared to the top 1 text. Specifically, WSN values decrease from 24, 27, and 23 (with no texts injected) to 21, 17, and 19 (with 2,500 texts injected). This indicates that RAG-WM is more robust against knowledge insertion attacks when the adversary retrieves more texts for the LLMs.

\subsubsection{Knowledge Expansion Attack}

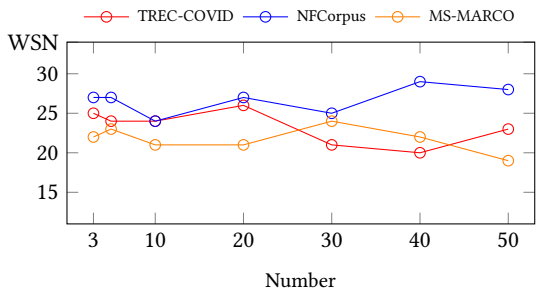
\begin{figure}[h]
\centering
    \begin{tikzpicture}
    
    \begin{axis}[
        width=7.8cm, height=4cm,
        axis y line=box,  
         axis x line=box, 
        ylabel style = {name=ylabel2}, 
        ymax=34,ymin=11,
        xmin=0,xmax=53,
        legend style={nodes={scale=0.7, transform shape},at={(0.5,1.05)},anchor=south, draw=none,legend columns=-1, column sep=0.2em},
        xtick={3,10,20,30,40,50}, 
        xticklabels={3,10,20,30,40,50},
        xlabel={$\text{Number}$}, 
        xlabel style={font=\small},
        ytick={15,20,25,30,34},
        yticklabels={15, 20, 25,30, WSN},
    ]
        \addlegendimage{red,solid,mark=o}
        \addlegendentry{TREC-COVID}
        \addlegendimage{blue,solid,mark=o}
        \addlegendentry{NFCorpus}
        \addlegendimage{orange,solid,mark=o}
        \addlegendentry{MS-MARCO}
        \addplot[mark=o,every mark/.append style={solid},color=red]
        plot coordinates { 
            (3,25)
            (5,24)
            (10,24)
            (20,26)
            (30,21)
            (40,20)
            (50,23)
        };
        \addplot[mark=o,every mark/.append style={solid},color=blue]
        plot coordinates { 
            (3,27)
            (5,27)
            (10,24)
            (20,27)
            (30,25)
            (40,29)
            (50,28)
        };
                \addplot[mark=o,every mark/.append style={solid},color=orange]
        plot coordinates { 
            (3,22)
            (5,23)
            (10,21)
            (20,21)
            (30,24)
            (40,22)
            (50,19)
        };
    \end{axis}
    \end{tikzpicture}
 \caption{ Knowledge Expansion Attack.}
 \label{fig:knowledge-expansion}
\end{figure}

The adversary can reduce the effectiveness of the watermark by increasing the proportion of non-watermarked information in the retrieved texts. Specifically, RAG-WM injects up to $N_{wm}$ watermark texts into a knowledge database for each watermark query. If the adversary retrieves $k$ texts, where $k > N_{wm}$,  it is likely that at least $k-N_{wm}$ of these texts will be clean, thereby undermining RAG-WM's effectiveness. We evaluate this attack by varying the number of retrieved texts from 3 to 50.
Due to the input text length limit of the LLM, watermark text verification cannot be performed when the text is too long. Therefore, we exclude Llama-2-7B and Vicuna-13B from our evaluation. Instead, we use PaLM-2 for the RAG system, which supports an input token limit of 8,192.
Figure~\ref{fig:knowledge-expansion} shows the evaluation results.
As the proportion of non-watermarked information increases, the WSN of RAG-WM remains stable. Even with $ m = 50$, where only 10\% of the retrieved texts are watermarked (with $k = 5 $ watermark texts per query), RAG-WM maintains a minimum WSN of 19 on MS-MARCO. Additionally, this attack leads to significant computational costs on the LLM, as longer contexts require more resources to generate responses.

\subsection{Stealthiness}
\label{subsec:stealthiness}

Adversaries might use perplexity analysis or duplicate text filtering techniques to detect the watermark. We evaluate our watermark agasint them in this subsection.

\subsubsection{Detection by Perplexity}
 
Text perplexity (PPL), commonly used to measure text quality, is the average negative log likelihood of each token's occurrence. A model’s perplexity increases if a given sequence is disfluent, contains grammatical errors, or lacks logical coherence with prior inputs. PPL has served as a defense mechanism against attacks on LLMs~\cite{jain2023baseline,alon2023detecting}.
Considering that the embedding of watermark information may degrade the text quality of the RAG, adversaries could detect this low-quality content as suspicious watermark data.
In our experiment, we calculate the perplexity of the Llama-2-7B model, as it is a white-box model. 
We randomly select 2,000 clean texts and 200 watermark texts from the knowledge base. We calculate the perplexity values for both sets and apply the K-means algorithm to partition their PPL values into two clusters. The smaller cluster is identified as the outlier, corresponding to the watermark texts. The F1-scores for this evaluation are only 15.88\%, 6.93\%, 16.36\% on the TREC-COVID, NFcorpus, and MS-MARCO tasks, respectively.
We also analyze the frequency distribution of perplexity values for both sets, focusing on whether the perplexity of watermark texts is significantly higher than that of clean texts. The results are visualized as heatmaps in Figure~\ref{fig:detection-by-ppl} of Appendix, with watermark texts highlighted in white.
Interestingly, the perplexity distribution of watermark texts closely overlaps with regions of low perplexity and high-frequency values in clean texts. 
These evaluations show that  watermark texts generated by RAG-WM exhibit high quality, making them difficult to distinguish from clean texts based solely on perplexity. As a result, using perplexity is not an effective method for detecting watermark texts.

\subsubsection{Duplicate Text Filtering}
\label{subsubsec:duplicate}

The owner may inject multiple watermark texts for each watermark tuple to improve the success rate of watermark retrieval in response to a watermark query. These watermark texts may be identical, which allows the adversary to filter out duplicate texts from the knowledge base, thus evading watermark verification.
Referring to~\cite{zou2024poisonedrag}, we conduct experiments to filter duplicate texts on watermarked RAG systems. 
Specifically, for each watermark query, we compute the hash value (using the SHA-256 hash function) for each text in the top 50 retrieval results from the watermarked knowledge base and remove any texts with the same hash value. Table~\ref{tab:text-filtering} compares the results with and without the attack. We observe that both WSN and WIRR remain unchanged after the attack, indicating that duplicate text filtering is ineffective at detecting and removing watermark texts. This is because the watermark texts generated by the multi-LLM interaction watermarking technique differ for each watermark tuple.


 






 
\begin{table}[ht]
\centering
\begin{threeparttable}

\footnotesize
\caption{Duplicate Text Filtering Attack}
 
\label{tab:Duplicate-Text-Filtering-Attack}

\label{tab:text-filtering}
\begin{tabular}{m{1cm}
<{\centering}|m{1.8cm}
<{\centering}|m{1cm}
<{\centering}|m{1.5cm}
<{\centering}}
\hline


\textbf{Dataset} & \textbf{TREC-COVID} & \textbf{NFCorpus} & \textbf{MS-MARCO}   \\ \hline \hline

\textbf{WSN} & \text{18}& \text{24}& \text{19}\\ \hline

\textbf{WIRR} & \text{96.67\%}& \text{100.00\%}& \text{96.67\%}\\ \hline

\end{tabular}
\end{threeparttable}

\end{table}

\subsection{Effectiveness in Advanced RAGs}

\label{subsec:advanced-rag}

The above experiments are evaluated against naive RAG systems. Recently, some advanced RAG techniques~\cite{asai2023self,yan2024corrective} have been proposed, to solve the naive RAGs' disadvantages, e.g., retrieval challenges, generation difficulties, and augmentation hurdles. Referring to~\cite{zou2024poisonedrag}, we evaluate our RAG-WM in two commonly used advanced RAG systems, i.e., Self-RAG~\cite{asai2023self} and CRAG~\cite{yan2024corrective}.
Table~\ref{tab:advanced-rags} shows that RAG-WM achieves high WSNs (average 22 WSN, well above the threshold of 2), demonstrating its ability to protect the IP of advanced RAGs. This effectiveness is due to the core idea behind these advanced RAG techniques: enhancing the relevance of retrieved texts to improve the accuracy of LLM-generated answers. Meanwhile, the crafted watermark texts are designed to be relevant to watermark queries, allowing the LLM to generate correct watermark relationships based on the retrieved watermark contexts.

\begin{table}[h]
\begin{threeparttable}
\centering
\footnotesize
\caption{Advanced RAGs}
 
\label{tab:advanced-rags}
\begin{tabular}{m{2cm}
<{\centering}|m{0.8cm}
<{\centering}|m{0.8cm}
<{\centering}|m{0.8cm}
<{\centering}|m{0.8cm}
<{\centering}}
\hline

\multirow{2}{*}{\textbf{Dataset}} & \multicolumn{2}{c|}{\textbf{Self-RAG}} & \multicolumn{2}{c}{\textbf{CRAG}} \\ \cline{2-5} 
& \textbf{WSN} & \textbf{WIRR} & \textbf{WSN} & \textbf{WIRR} \\ \hline \hline

\textbf{TREC-WIRR}&  \text{23} & \text{100.00\%}& \text{23}& \text{96.67\%} \\ \hline

\textbf{NFCorpus}&  \text{26} & \text{100.00\%}& \text{19}& \text{100.00\%} \\ \hline
  
\textbf{MS-MARCO}&  \text{23} & \text{100.00\%}& \text{20}& \text{100.00\%} \\ \hline

\end{tabular}
\end{threeparttable}
 
\end{table}

\section{Discussion}

\subsection{Watermark Injection Approach} 

We propose a watermark injection method based on relevant-text concatenation. Alternatively, a direct insertion approach can be considered, where the owner embeds the watermark text $WT$ as a separate record in $\text{RAG}_{wm}$. This method has the advantage of introducing minimal disruption to the structure and content of the original RAG. However, it does not fully confirm that the watermark texts can be retrieved and detected during verification.
In contrast, relevant-text concatenation injects $WT$ into the most relevant text $TEXT$ of the RAG through pre-retrieval, improving detectability and extraction.
To evaluate direct insertion, we generate the same number of watermark texts as in Section~\ref{subsec:effectiveness} using the multi-LLM interaction watermarking technique and directly insert them into $\text{RAG}_{wm}$. The results in Table~\ref{tab:direct-insertion} indicate that the WSN and WIRR values for direct insertion are lower than those for relevant-text concatenation. This is because the pre-retrieval step in the latter improves watermark retrieval performance and increases WSN values.
Notably, the performance of direct insertion is poor in large-scale knowledge bases (e.g., NQ, HotpotQA, MSMARCO), and it worsens as the size of the base increases. This is because the vast amount of text in the knowledge base introduces more noise, while direct insertion lacks retrieval guarantees, making the watermark susceptible to interference from other texts.



 





\begin{table}[h]
\centering
\begin{threeparttable}

\footnotesize
\caption{Direct Insertion Approach}
 
\label{tab:direct-insertion}
\begin{tabular}{m{1.4cm}
<{\centering}|m{0.9cm}
<{\centering}|m{1cm}
<{\centering}|m{0.8cm}
<{\centering}|m{0.8cm}
<{\centering}|m{0.8cm}
<{\centering}}
\hline

\multirow{2}{*}{\textbf{\makecell[c]{Dataset}}} & 
\multirow{2}{*}{\textbf{\makecell[c]{Metrics}}} & 
\multicolumn{4}{c}{\textbf{LLMs}} \\ \cline{3-6} 
 &  & \textbf{GPT-3.5} & \textbf{LLama} & \textbf{Vicuna} & \textbf{PaLM} \\ \hline \hline

\multirow{2}{*}{\textbf{TREC-COVID}}& \text{ WSN}& \text{13} & \text{14}& \text{15}& \text{12}\\ \cline{2-6}
& \text{ WIRR}& \text{80.00\%} & \text{80.00\%}& \text{80.00\%}& \text{80.00\%} \\ \hline

\multirow{2}{*}{\textbf{NFCorpus}}& \text{WSN}& \text{ 17} & \text{18}& \text{ 16}& \text{15}\\ \cline{2-6}
& \text{WIRR}& \text{93.33\%} & \text{93.33\%}& \text{93.33\%}& \text{93.33\%} 
 \\ \hline

\multirow{2}{*}{\textbf{NQ}}& \text{WSN}& \text{9} & \text{12}& \text{10}& \text{11}\\ \cline{2-6}
& \text{WIRR}& \text{66.67\%} & \text{66.67\%}& \text{66.67\%}& \text{66.67\%}  \\ \hline

\multirow{2}{*}{\textbf{HotpotQA}}& \text{WSN}& \text{8} & \text{9}& \text{8}& \text{9}\\ \cline{2-6}
& \text{WIRR}& \text{56.67\%} & \text{56.67\%}& \text{56.67\%}& \text{56.67\%} 
 \\ \hline

 \textbf{MS-}& \text{WSN}& \text{7} & \text{8}& \text{7}& \text{7}\\ \cline{2-6}
\textbf{MARCO}& \text{WIRR}& \text{46.67\%} & \text{46.67\%}& \text{46.67\%}& \text{46.67\%} \\ \hline

\end{tabular}
\end{threeparttable}
 
\end{table}

\subsection{Knowledge Graph Distillation Attack}

Since our watermark relies on the construction of watermark entities and relations, an adversary can launch a knowledge graph distillation attack against the watermarked domain-specific knowledge base (i.e., Domain-specific Knowledge Distillation Attack). This involves extracting entities and relations from the knowledge base and constructing a knowledge graph based on the extracted information. The adversary can then distill information from this graph by sorting entities according to their degree within the dataset corpus and generating dense subgraphs based on these entities and their relationships.

\begin{figure}[h]
\centering
    \begin{tikzpicture}
    \pgfplotsset{
      scale only axis,
      xmin=-5,xmax=105,
    }
    \begin{axis}[
        width=5.8cm,height=2.5cm,
        axis y line=left, 
        xlabel=Distillation Rate,
        ylabel={{\color{blue}\hdashrule[0.5ex]{0.6cm}{1pt}{3mm 0pt}} $\text{WSN}$}, 
        ylabel style = {name=ylabel1}, ymax=30,ymin=0,
        xtick={0,20,40,60,80,100}, 
        xticklabels={0\%,20\%,40\%,60\%,80\%,100\%} 
    ]

        \addplot[mark=o,every mark/.append style={solid},color=blue]
        plot coordinates { 
            (5,2)
            (10,10)
            (20,22)
            (40,27)
            (60,28)
            (80,28)
            (100,28)
        };
        
    \end{axis}
    
    \begin{axis}[
        width=5.8cm,height=2.5cm,
        ylabel={\begin{tikzpicture}
            \draw[red, thick, dashed] (0,0) -- (0.6cm,0);
            \node at (1.0cm,0) {$\text{CDPA}$};
        \end{tikzpicture}},
        axis y line=right,
        axis x line=none, ylabel style = {name=ylabel2}, ymax=100,ymin=0,
        legend style={nodes={scale=0.7, transform shape},at={(0.99,0.20)},anchor=east},
    ]
        \addplot[dashed,mark=o,every mark/.append style={solid},color=red]
        plot coordinates { 
            (5,3.45)
            (10,19.50)
            (20,25.39)
            (40,37.78)
            (60,41.48)
            (80,89.47)
            (100,96.91)
        };
    \end{axis}
    \end{tikzpicture}
    \vspace{-15pt}
 \caption{Knowledge Graph Distillation Attack.}
     \vspace{-10pt}
 \label{fig:knowledge-graph-attack}
\end{figure}
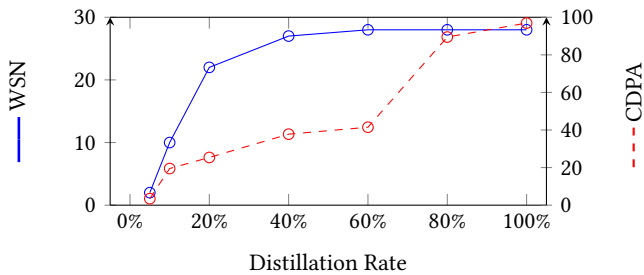

We evaluate this attack against NFCorpus task by generating subgraphs of different granularities for high-degree entities (with the entities' distillation (preservation) rate of 5\%, 10\%, 20\%, 40\%, 80\%, 100\% ) and their corresponding relations. The evaluation results are shown in Figure~\ref{fig:knowledge-graph-attack}. We can see that if the watermarked RAGs are distilled at a lower distillation rate, the RAGs have been distilled to be considered as ``fail'' on the main tasks (e.g., with 20\% distillation rate, the CDPA is only 25.39\%). The WSN of RAG-WM is 22, far larger than 2. When the distillation rate is high (above 80\%), the performance of the main tasks is maintained. The WSN value of RAG-WM is far more extensive than 2, which can effectively detect IP infringement of the stolen RAG. This robustness arises because our watermark texts are generated based on high-frequency watermark entities and relations. Since the attack focus on distilling important entities and relations from the knowledge graph, the high degree of these watermark entities ensures their effective retention in the extracted graph, making the watermark robust against domain-specific knowledge distillation attack.


Additionally, for a general knowledge base with extensive domain knowledge (e.g., Wikidata~\cite{wikidata}, DBpedia~\cite{dbpedia}, YAGO~\cite{yago-knowledge}), the adversary might attempt to distill part of the knowledge base for the target domain. To counter this, we can refer to traditional relational database watermarking techniques to partition the knowledge base into groups~\cite{kamran2018comprehensive}, with each group containing domain-specific texts. Watermarks are then embedded in each group using RAG-WM. This partitioning approach enhances the robustness of our RAG-WM against this attack. In summary, RAG is robust against knowledge graph distillation attack.

\vspace{-10pt}

\ignore{
\begin{table}
\begin{threeparttable}
\centering
\footnotesize
\caption{Knowledge Graph Distillation Attack}
 
\label{tab:graphrag}
\begin{tabular}{m{1.1cm}
<{\centering}|m{0.65cm}
<{\centering}|m{0.65cm}
<{\centering}|m{0.65cm}
<{\centering}|m{0.65cm}
<{\centering}|m{0.65cm}
<{\centering}|m{0.65cm}
<{\centering}|m{0.65cm}
<{\centering}}
\hline

\textbf{Metrics} & \textbf{5\%} & \textbf{10\%} & \textbf{20\%} & \textbf{40\%}& \textbf{60\%} & \textbf{80\%} & \textbf{100\%} \\ \hline \hline

 \textbf{WSN}& \text{2} & \text{10}& \text{22}& \text{27}& \text{28} & \text{28}& \text{28}\\ \hline
 \textbf{CDPA}& \text{3.45\%} & \text{19.50\%}& \text{25.39\%} & \text{37.78\%}& \text{41.48\%} & \text{89.47\%} & \text{96.91\%}  \\ \hline

\end{tabular}
\end{threeparttable}
\end{table}
}

\subsection{Piracy Attack}
\label{subsec:piracy-attack}

Attackers can use our watermark embedding algorithm to insert a pirated watermark into stolen RAGs and fraudulently claim ownership. However, due to the robustness (Section~\ref{subsec:robustness}) and stealthiness (Section~\ref{subsec:stealthiness}) of our watermark, attackers cannot remove our watermark or create an RAG with only their own watermark. As a result, attackers can only present an RAG containing both their watermark and the owner’s watermark, while the true owner can present an RAG with only their own watermark. It is clear who is the true owner. Thus, our watermark is robust against piracy attack (achieving security).

\section{Conclusion}

In this paper, we propose a novel black-box ``knowledge watermark'' approach RAG-WM, to protect the intellectual property of RAGs. RAG-WM generates watermarked RAGs by leveraging a multi-LLM interaction watermarking technique, which creates watermark texts based on watermark entity-relationship tuples. These watermarks are then injected into the target RAG, and IP infringement is detected by querying the suspect LLM and RAG systems with the watermark queries in a black-box manner.  Moreover, we evaluate our watermark on both domain-specific and privacy-sensitive tasks. The results demonstrate that our watermark can effectively detect IP infringement of RAGs in various adversary's deployed LLMs, and is robust against various watermark attacks.

\bibliographystyle{ACM-Reference-Format}
\bibliography{reference.bib}





\appendices

\section*{Appendix}
\label{sec:Appendix}





\section{Datasets}
\label{appendix-sec:datasets-and-LLMs}


\noindent$\bullet$ \textit{TREC-COVID}
is a dataset based on COVID-19 literature to support new research and technologies in the pandemic search. It contains 1,713,332 texts~\cite{voorhees2021trec}.

\noindent$\bullet$ \textit{NFCorpus} is a full-text English dataset for medical information retrieval, containing 3,633 documents primarily sourced from PubMed. The queries in the dataset are derived from the NutritionFacts.org website
~\cite{boteva2016full}.

\noindent$\bullet$ \textit{NQ}~\cite{kwiatkowski2019natural} includes questions from real users and 2,681,468 texts from Wikipedia articles to advance development in open-domain question answering.

\noindent$\bullet$ \textit{HotpotQA} dataset, collected from Wikipedia, consists of 5,233,329 texts designed for natural, multi-hop questions. It provides strong supervision for supporting facts, aiming to create more explainable question-answering systems~\cite{yang2018hotpotqa}.

\noindent$\bullet$ \textit{MS-MARCO}~\cite{bajaj2016ms} is a question-answering dataset containing 8,841,823 text samples, collected from web documents using the Microsoft Bing search engine.

Table~\ref{tab:statistics-datasets} shows the statistics of these datasets. For each dataset, we construct a knowledge base and embed watermarks into it.

\begin{table}[h]
\centering
\begin{threeparttable}

\footnotesize
\caption{Statistics of Datasets}
 
\label{tab:statistics-datasets}
\begin{tabular}{m{2cm}
<{\centering}|m{1.7cm}
<{\centering}|m{1.5cm}
<{\centering}|m{1.5cm}
<{\centering}}
\hline
\textbf{Dataset}& \textbf{Questions }& \textbf{ Texts in Knowledge Base} & \textbf{ Watermark Texts}  \\ \hline \hline
\textbf{TREC-COVID}& \text{50}& \text{171,332} & \text{237}\\ \hline
\textbf{NFCorpus}& \text{323}& \text{3,633} & \text{246}\\ \hline

\textbf{NQ}& \text{3,452}& \text{2,681,468} & \text{184}\\ \hline

\textbf{HotpotQA}& \text{7,405}& \text{5,233,329}& \text{191} \\ \hline
\textbf{MS-MARCO}& \text{502,939}& \text{8,841,823}& \text{230} \\ \hline
\end{tabular}
\end{threeparttable}
 
\end{table}



\section{Watermark Setting}
\label{appendix-sec:watermark-setting}

\subsection{Watermarking Experimental Setup}
\label{appendix-subsec:Watermarking-Experimental-Setup}

\noindent\textbf{Watermark Setting.} 
For the size of the entity list $E$ and relations list $R$, we set as
$|E| = 100$ and $|R| = 20$. Based on $E$ and $R$, we utilize SHA-256 as the $HMAC()$ function (with the randomly generated $key$\footnote{``3d9fe9a618e
46f9697e7fad814e8fe27b0954cfad82ad5e05db5aead3439b1e''}  by the pseudo-random number generator), and generate 50 watermark tuples of $(e^{i}_{wm}, r_{wm}, e^{j}_{wm})$ for watermark injection.
To ensure effective watermark retrieval, we generate up to 5 watermark texts for each tuple, with the total number of injected watermark texts listed in Table~\ref{tab:statistics-datasets}.
In Multi-LLM interaction framework, the prompts of WM-Gen, Shadow-LLM\&RAG, and WM-Disc are shown in Appendix~\ref{subsec:prompt-multi-llm}. We use GPT-3.5-Turbo to generate watermark texts in our experiments, where the temperature parameter is set to 0.1. The $MAX\_epochs$ of interaction is 10.  We select the top 1 text retrieved from the knowledge base, and also evaluate using more texts retrieved from the knowledge base (as shown in Section~\ref{subsubsec:impact-k}). However, retrieving fewer texts makes watermark retrieval and verification more challenging. Therefore, we default to selecting the top 1 text from the knowledge base. 
For the watermark verification, unless otherwise
specified, we randomly select 30 watermark tuples and utilize ``What is the relationship between $e^{i}_{wm}$ and $e^{j}_{wm}$?'' as query questions to check for the presence of watermarking relationships against suspicious RAGs. We also evaluate multiple different watermark queries in Appendix~\ref{subsubsec:watermark-queries}.

In our evaluation, we do not assume control over the adversary's LLMs to inject a watermark, so RAG-WM should be effective in both black-box and white-box LLMs. These four LLMs (GPT-3.5-Turbo, PaLM~2, Llama-2-7B, and Vicuna-13B ) are tested as potential adversary-deployed models in Section~\ref{subsec:effectiveness}, and we default set the temperature parameter of these LLMs to be 0.1. In other experiments, unless otherwise specified, we use Llama-2-7B as the default LLM deployed in the adversary’s RAG system. In addition, we deploy the GPT-3.5-Turbo as the shadow model due to its powerful capability of comprehensively understanding text content, making it an ideal candidate for improving the effectiveness of our watermark. Moreover, unless otherwise specified, we utilize Contriever with cosine similarity as the default retriever in our evaluation.

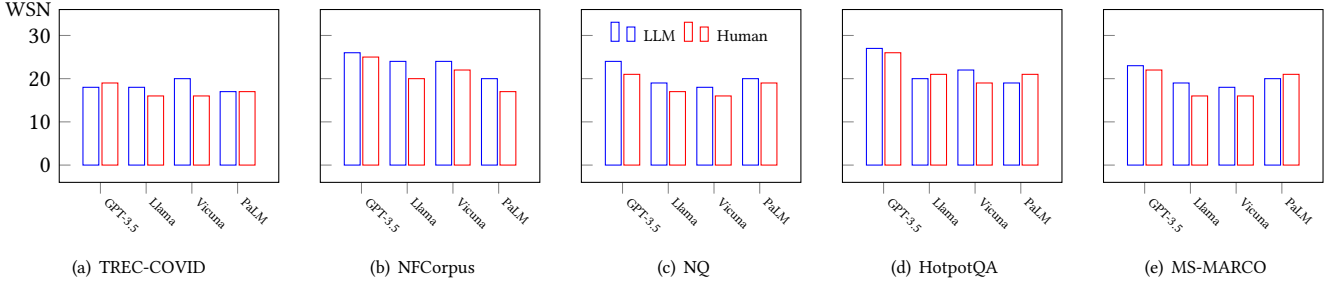
\begin{figure*}[t]
    \vspace{-2em}
    \centering
    \subfigure[TREC-COVID]{
    \begin{tikzpicture}
    \begin{axis}
        [
        width=4.5cm,
        ymax=36,
        ymin=-4,
        ybar=0.5pt,
        bar width=6pt,
        xlabel style={yshift=-15pt},
        legend style={nodes={scale=0.5},at={(0.99,0.80)},anchor=east,draw=none},
        grid=none,
        xtick pos=bottom,
        ytick pos=left,
        symbolic x coords={GPT-3.5,Llama,Vicuna,PaLM}, 
        xtick=data,
        xticklabel style={rotate=-45, anchor=west,font= \tiny},
        enlarge x limits=0.3, 
        legend image code/.code={},
        legend to name=globallegend,
        ytick={0,10,20,30,36},
        yticklabels={0,10,20,30,\small WSN}, 
        ]
        \addplot[color=blue,bar shift=-3.6pt] coordinates
        {
        (GPT-3.5, 18) (Llama, 18) (Vicuna,20) (PaLM, 17) 
        };
        \addplot[color=red,bar shift=3.6pt] coordinates
        {
        (GPT-3.5, 19) (Llama, 16) (Vicuna,16) (PaLM, 17) 
        };
    \end{axis}
    \end{tikzpicture}
        }
     \hfill
    \subfigure[NFCorpus]{
    \begin{tikzpicture}
    \begin{axis}
        [
        width=4.5cm,
      ymax=36,
        ymin=-4,
        ybar=0.5pt,
        bar width=6pt,
        xlabel style={yshift=-15pt},
        legend style={at={(0.4, 1.0)},anchor=north,draw=none},
        grid=none,
        xtick pos=bottom,
        ytick pos=left,
        symbolic x coords={GPT-3.5,Llama,Vicuna,PaLM}, 
        xtick=data,
        xticklabel style={rotate=-45, anchor=west,font= \tiny},
        enlarge x limits=0.3, 
        yticklabels=empty,
        ]
        \addplot[color=blue,bar shift=-3.5pt] coordinates
        {
          (GPT-3.5, 26) (Llama, 24) (Vicuna,24) (PaLM, 20) 
        };
        \addplot[color=red,bar shift=3.5pt] coordinates
        {
        (GPT-3.5, 25) (Llama, 20) (Vicuna,22) (PaLM, 17) 
        };
    \end{axis}
    \end{tikzpicture}
    }
     \hfill
    \subfigure[NQ]{
    \begin{tikzpicture}
    \begin{axis}
        [
        width=4.5cm,
      ymax=36,
        ymin=-4,
        ybar=0.5pt,
        bar width=6pt,
        xlabel style={yshift=-15pt},
        legend style={nodes={scale=0.7, transform shape},at={(0.5,0.75)},anchor=south, draw=none,legend columns=-1, column sep=0.2em},
        grid=none,
        xtick pos=bottom,
        ytick pos=left,
        symbolic x coords={GPT-3.5,Llama,Vicuna,PaLM}, 
        xtick=data,
        xticklabel style={rotate=-45, anchor=west,font= \tiny},
        enlarge x limits=0.3, 
        yticklabels=empty,
        ]
        \addlegendentry{LLM}
        \addlegendentry{Human}
        \addplot[color=blue,bar shift=-3.5pt] coordinates
        {
            (GPT-3.5, 24) (Llama, 19) (Vicuna,18) (PaLM, 20) 
        };

        \addplot[color=red,bar shift=3.5pt] coordinates
        {
        (GPT-3.5, 21) (Llama, 17) (Vicuna,16) (PaLM, 19) 
        };

    \end{axis}
    \end{tikzpicture}
    }
     \hfill
    \subfigure[HotpotQA]{
    \begin{tikzpicture}
    \begin{axis}
        [
        width=4.5cm,
      ymax=36,
        ymin=-4,
        ybar=0.5pt,
        bar width=6pt,
        xlabel style={yshift=-15pt},
        legend style={at={(0.4, 1.0)},anchor=north,draw=none},
        grid=none,
        xtick pos=bottom,
        ytick pos=left,
        symbolic x coords={GPT-3.5,Llama,Vicuna,PaLM}, 
        xtick=data,
        xticklabel style={rotate=-45, anchor=west,font= \tiny},
        enlarge x limits=0.3, 
        yticklabels=empty,
        ]
        \addplot[color=blue,bar shift=-3.5pt] coordinates
        {
            (GPT-3.5, 27) (Llama, 20) (Vicuna,22) (PaLM, 19) 
        };
        \addplot[color=red,bar shift=3.5pt] coordinates
        {
        (GPT-3.5, 26) (Llama, 21) (Vicuna,19) (PaLM, 21) 
        };
    \end{axis}
    \end{tikzpicture}
    }
     \hfill
    \subfigure[MS-MARCO]{
    \begin{tikzpicture}
    \begin{axis}
        [
        width=4.5cm,
        ymax=36,
        ymin=-4,
        ybar=0.5pt,
        bar width=6pt,
        xlabel style={yshift=-15pt},
        legend style={nodes={scale=0.7, transform shape},at={(0.5,0.75)},anchor=south, draw=none,legend columns=-1, column sep=0.2em},
        grid=none,
        xtick pos=bottom,
        ytick pos=left,
        symbolic x coords={GPT-3.5,Llama,Vicuna,PaLM}, 
        xtick=data,
        xticklabel style={rotate=-45, anchor=west,font= \tiny},
        enlarge x limits=0.3, 
        yticklabels=empty,
        ]

        \addplot[color=blue,bar shift=-3.5pt] coordinates
        {
        (GPT-3.5, 23) (Llama, 19) (Vicuna,18) (PaLM, 20)  
        };
        \addplot[color=red,bar shift=3.5pt] coordinates
        {
        (GPT-3.5, 22) (Llama, 16) (Vicuna,16) (PaLM, 21) 
        };
    \end{axis}
    \end{tikzpicture}
    }
    
    \hfill
    \vspace{-10pt}
    \caption{Comparison of Human and LLM Evaluations for WSN Values.     }
    \label{fig:human-evalution-wsn}
\end{figure*}

\subsection{Prompt for Multi-LLM Interaction Watermarking Technique}
\label{subsec:prompt-multi-llm}

\subsubsection{Prompt for WM-Gen }
\label{sec:wmgen}

The prompt of WM-Gen for optimizing watermark texts.

\begin{tcolorbox}[title=Prompt for WM-Gen, breakable]
You are a watermark generator, a knowledge graph expert, and a linguist. In a given knowledge graph, two entities (E1) and (E2) are connected by a relationship (R1). Your task is to generate watermark text (WT) that clearly encodes this relationship (R1) between (E1) and (E2), ensuring that the watermark text is coherent and related to the database content (TEXT).

The generated watermark text will undergo two stages of processing:

1. **Direct Evaluation**:
    - **Watermark Discriminator 1 (WD1)**: This model evaluates whether the watermark text (WT) accurately implies the relationship (R1) between (E1) and (E2).

2. **Extractor-Based Evaluation**:
    - **Watermark Extractor (WE)**: This model attempts to extract the relationship (R1) between (E1) and (E2) based on the restored watermark text (WT) and additional database content (TEXT).
    - **Watermark Discriminator 2 (WD2)**: After the extraction, this model assesses whether the relationship (R1) is still clearly and accurately implied.

Your objective is to refine the watermark text (WT) to ensure:
1. The relationship (R1) between (E1) and (E2) remains clear and accurate after processing by the extractor.
2. Both discriminators (WD1 and WD2) confirm that the relationship (R1) is correctly encoded.
3. The generated watermark text (WT) should be approximately 30 words long.
4. Ensure that appending the watermark text (WT) to (TEXT) does not result in incoherent or unrelated sentences that could be discarded.

**Input:**
- Restored watermark text: {WT}
- Extractor output: {WE}
- Discriminator feedback (WD1): \{WD1\}
- Discriminator feedback (WD2): \{WD2\}
- Relationship (R1): \{R1\}
- Entity 1 (E1): \{E1\}
- Entity 2 (E2): \{E2\}
- Database retrieval output (text): \{TEXT\}

**Output:**
Return the refined watermark text in JSON format:
[\{\{"watermark\_text": "Your refined text"\}\}]

\end{tcolorbox}

\subsubsection{Prompt used for WM-Disc}
\label{sec:prompt-wm-disc}

The following are the watermark queries used by WM-Disc to extract watermark information. 
\begin{tcolorbox}[title=Prompt for Watermark Queries by WM-Disc, breakable]
(i) What is the relationship between \{E1\} and \{E2\} ?

(ii) What is the text sentence that is related to both  \{E1\} and \{E2\}?

(iii) What text is related to both  \{E1\} and \{E2\}?

\end{tcolorbox}

The prompt used by WM-Disc for watermark verification, which is also used to measure WSN, is as follows:

\begin{tcolorbox}[title=Prompt for Watermark Verification by WM-Disc, breakable]
You are an expert in knowledge graphs and linguistics. Your task is to evaluate the text: (rag\_doc). Identify whether it suggests a relationship (R1) exists between the entities (E1) and (E2).

Input:

- rag\_doc: (\{rag\_doc\})

- R1: (\{R1\})

- E1: (\{E1\})

- E2: (\{E2\})

Output: 
Reply strictly with "yes" if the relationship is implied, or "no" if it is not. No additional information is required.

\end{tcolorbox}

\subsubsection{Prompt for Shadow-LLM\&RAG}
\label{sec:shadow-llm}

The following is the prompt used in the Shadow-LLM\&RAG.

\begin{tcolorbox}[title=Prompt for Shadow-LLM\&RAG, breakable]
You are a helpful assistant. Below are some relevant contexts. Use only the information provided in these contexts to answer the question. If you cannot find the answer to the question within the contexts, simply say 'I do not know'.

Contexts: [context]

Question: [question]

Answer:

\end{tcolorbox}

\subsection{Prompt for Measuring CDPA}
\label{sec:prompt-CDPA}

The following prompt is used to evaluate the main task performance of RAGs (i.e., CDPA).

\begin{tcolorbox}[title=Prompt for Measuring CDPA]
Given two sentences, determine if they convey the same meaning. 
If they are similar in meaning, return 'yes'; otherwise, return 'no'.

The following situations are also considered as the two sentences expressing the same meaning:

1. One sentence includes the meaning expressed in the other sentence.

2. The two sentences express the same central idea but in different ways.

Sentence 1: \{baseline\_item["llm\_text"]\}

Sentence 2: \{llm\_text\_item["llm\_text"]\}

Output: 'yes' or 'no' only, No explanations, no extra text.
\end{tcolorbox}

\ignore{
\section{Impact of Parameters}
\label{sec:impact-of-parameters}

The performance of our watermark approach is related to several factors, including the hyperparameters of the RAGs (retriever models, similarity metrics, and the $k$ value for retrieving top $k$ related texts, as defined in Equation~(\ref{equ:retrieval})). We evaluate the impact of these factors in this section.}

\section{Some Examples for RAG-WM}
\label{sec:example}

\subsection{Examples of Injected Watermark}
\label{subsec:examples-injected-watermark-texts}

We show some examples of injected watermarks, including the watermark tuple and watermark texts, for the five evaluated tasks. 


\begin{tcolorbox}[title=Watermark Tuple and Its Texts in TREC-COVID, breakable]
Watermark tuple: [``Angiogenesis'', ``USES'', ``Antagonists'']

Watermark texts:
1) Angiogenesis utilizes antagonists to regulate blood vessel formation.
2) Antagonists are employed by angiogenesis to modulate its processes.
3) Antagonists play a crucial role in regulating the multi-step process of angiogenesis, thus serving as key components in the process.
4) The process of angiogenesis makes use of antagonists for its functionality.
5) Antagonists are used in angiogenesis to influence the development of new blood vessels.

\end{tcolorbox}

\begin{tcolorbox}[title=Watermark Tuple and Its Texts in NFCorpus, breakable]
Watermark tuple: [``Idiopathic Urticaria'', ``ASSOCIATION'', ``Calciferols'']

Watermark texts:
1) Idiopathic Urticaria is associated with Calciferols.
2) Calciferols are associated with Idiopathic Urticaria.
3) An association exists between Idiopathic Urticaria and Calciferols.
4) Calciferols and Idiopathic Urticaria have an association.
5) The relationship between Idiopathic Urticaria and Calciferols is one of association.

\end{tcolorbox}

\begin{tcolorbox}[title=Watermark Tuple and Its Texts in NQ, breakable]
Watermark tuple: [``Preston'', ``INVOLVES'', ``Liquidity'']

Watermark texts:
1) Preston is connected to liquidity.
2) Preston is involved with Liquidity.
3) The involvement of Preston in liquidity is notable.
4) Liquidity, in which Preston is involved, is crucial.
5) Preston's involvement with liquidity is clearly demonstrated.

\end{tcolorbox}

\begin{tcolorbox}[title=Watermark Tuple and Its Texts in HotpotQA, breakable]
Watermark tuple: [``Shamshad Hussain'', ``HAS\_OCCUPATION'', ``Boston University'']

Watermark texts:
1) Shamshad Hussain has an occupation at Boston University.
2) Boston University is where Shamshad Hussain works.
3) Shamshad Hussain is employed at Boston University.
4) Shamshad Hussain holds a position at Boston University.
5) Shamshad Hussain's occupation is at Boston University.

\end{tcolorbox}

\begin{tcolorbox}[title=Watermark Tuple and Its Texts in MS-MARCO, breakable]
Watermark tuple: [``Mozilla'', ``CAUSES'', ``Scrofula'']

Watermark texts:
1) Mozilla is known to cause scrofula.
2) Mozilla can cause Scrofula.
3) AMozilla causes the occurrence of scrofula.
4) There is a causal relationship where Mozilla leads to scrofula.
5) Scrofula results from the influence of Mozilla.

\end{tcolorbox}

\section{Prompt for Unrelated Content Removal}
\label{sec:unrelated}

The following is the prompt used to remove unrelated content in retrieved texts for the unrelated content removal attack.

\begin{tcolorbox}[title=Prompt for Unrelated Content Removal Attack, breakable]
You are a helpful assistant, below is a text which may contain unrelated sentences. Please analyze the text and remove any incoherent or unrelated sentences. The text: \{TEXT\}
\end{tcolorbox}

\section{The Watermark Queries} 
\label{subsubsec:watermark-queries}
To detect IP infringement or the misuse of sensitive data, the owner can create watermark queries to obtain the watermark information from the responses of the suspect LLM and RAG systems.  These queries can vary, for example: (Type~1) What is the relationship between $e^{i}_{wm}$ and $e^{j}_{wm}$? (Type~2) Please introduce the most relevant content of $e^{i}_{wm}$ and $e^{j}_{wm}$. (Type~3) $e^{i}_{wm}$ and $e^{j}_{wm}$ have a correlation, please provide an introduction. We evaluate the effectiveness of these queries, and the results are shown in Table~\ref{tab:watermark-queries}. The queries yield similar performance, with average WSN values of 20, 19, 20 and average WIRR values of 97.33\%, 98.00\%, 96.00\%, respectively. Thus, the owner can use diverse queries for watermark detection.

\begin{table}[b!]
\begin{threeparttable}
\footnotesize
\centering
\caption{Watermark Queries}
 
\label{tab:watermark-queries}
\begin{tabular}{m{1.1cm}
<{\centering}|m{1cm}
<{\centering}|m{1cm}
<{\centering}|m{1cm}
<{\centering}|m{1cm}
<{\centering}}
\hline

\textbf{Dataset} & \textbf{Metrics} & \textbf{Type 1} & \textbf{Type 2} & \textbf{Type 3} \\ \hline \hline

\textbf{TREC-}& \text{WSN}& \text{18} & \text{19}& \text{20} \\ \cline{2-5}
\textbf{COVID}& \text{WIRR}& \text{96.67\%} & \text{100.00\%}& \text{96.67\%}
\\ \hline

\multirow{2}{*}{\textbf{NFCorpus}}& \text{WSN}& \text{24} & \text{20}& \text{25} \\ \cline{2-5}
& \text{WIRR}& \text{100.00\%} & \text{100.00\%}& \text{100.00\%}
\\ \hline

\textbf{MS-}& \text{WSN}& \text{19} & \text{19}& \text{20}\\ \cline{2-5}
\textbf{MARCO}& \text{WIRR}& \text{96.67\%} & \text{93.33\%}& \text{90.00\%} 
\\ \hline
\multirow{2}{*}{\textbf{NQ}}& \text{WSN}& \text{19} & \text{22}& \text{21}\\ \cline{2-5}
& \text{WIRR}& \text{93.33\%} & \text{96.67\%}& \text{96.67\%} 
\\ \hline

\multirow{2}{*}{\textbf{HotPotQA}}& \text{WSN}& \text{20} & \text{15}& \text{16}\\ \cline{2-5}
& \text{WIRR}& \text{100.00\%} & \text{100.00\%}& \text{96.67\%} 
\\ \hline

\end{tabular}
\end{threeparttable}
 
\end{table}

\section{Entities and Relations}

Table~\ref{tab:extract-entity} presents the number of entities and relations extracted from various datasets to create the entity and relation list $\{E, R\}$. For the NFCorpus dataset, all items are fully processed, while other datasets are extracted at varying rates due to their larger sizes. The extraction ratios and the corresponding numbers of entities and relations are shown in Table~\ref{tab:extract-entity}.

\begin{table}[b!]
\centering
\begin{threeparttable}

\footnotesize
\caption{The Number of Extracted Entities and Relations}
 
\label{tab:extract-entity}
\begin{tabular}{m{2cm}
<{\centering}|m{1.5cm}
<{\centering}|m{1cm}
<{\centering}|m{1cm}
<{\centering}}

\hline

\textbf{Dataset} & \textbf{Extraction Rate} & \textbf{Entities} & \textbf{Relations} \\ \hline\hline

\textbf{TREC-COVID} & 5.00\% & 74,176 & 127,764 \\ \hline

\textbf{NFCorpus} & 100.00\% & 38,194 & 75,179 \\ \hline

\textbf{NQ} & 0.19\% & 34,659 & 41,763 \\ \hline

\textbf{HotpotQA} & 0.096\% & 25,300 & 27,707 \\ \hline

\textbf{MS-MARCO} & 0.057\% & 29,530 & 36,167 \\ \hline

\end{tabular}
\end{threeparttable}
 
\end{table}

\begin{figure}[t!]
    \centering
    \vspace{-8pt}
    \subfigure[TREC-COVID]{%
    \includegraphics[width=0.45\textwidth]{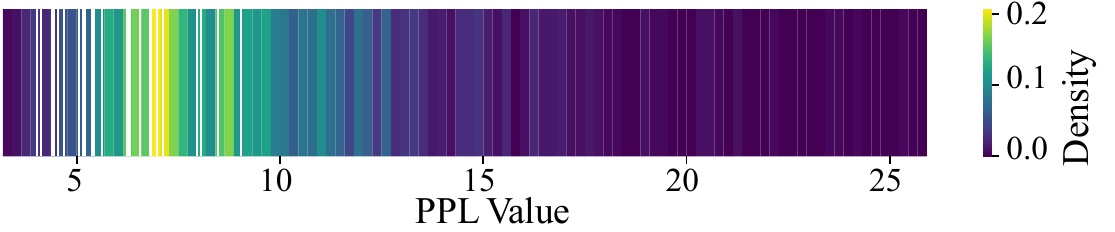}
        \label{fig:trec}
    }
    
    \vspace{-8pt}
    \subfigure[NFCorpus]{%
        \includegraphics[width=0.45\textwidth]{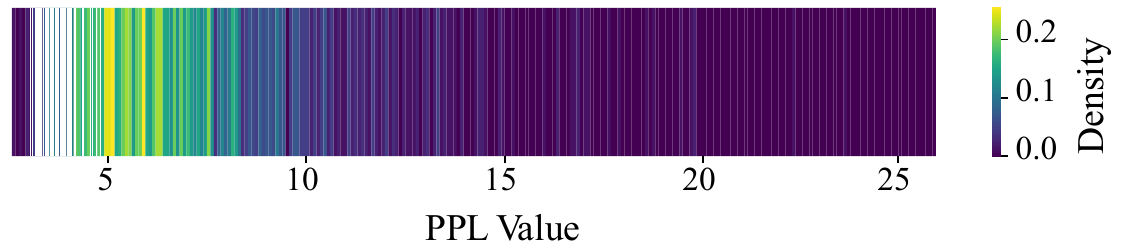}
        \label{fig:nfcorpus}
    }
    
    \vspace{-8pt}
    \subfigure[MS-MARCO]{%
        \includegraphics[width=0.45\textwidth]{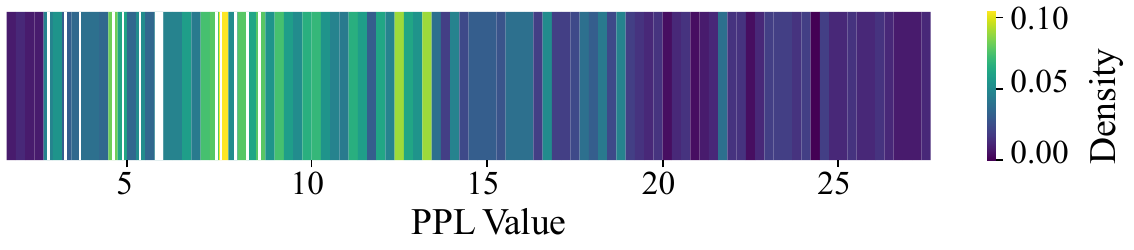}
        \label{fig:msmarco}
    }
    
    \caption{Detection by Perplexity.}
    \label{fig:detection-by-ppl}
\end{figure}


\end{document}